\documentclass[11pt]{article}

\usepackage[margin=1in]{geometry}
\usepackage{graphicx}
\usepackage[T1]{fontenc}
\usepackage[utf8]{inputenc}
\usepackage{times}
\usepackage{amsmath,amssymb}
\usepackage{booktabs}
\usepackage{setspace}
\usepackage{caption}
\renewcommand{\figurename}{Fig.}

\usepackage[numbers,sort&compress]{natbib}
\usepackage[colorlinks=true,linkcolor=blue,citecolor=blue,urlcolor=blue]{hyperref}
\usepackage{xcolor}

\graphicspath{{figures/}}
\title{A subsurface array of photonic crystal slabs produces green stripes in a scarab beetle}
\author{}
\date{}

\begin{document}

\begin{center}
{\LARGE\bfseries A subsurface array of photonic crystal slabs produces green stripes in a scarab beetle}\\[1.25em]
\large
Laura Ospina-Rozo,$^{1\ast\dag}$
Nicola S. Kubzdela,$^{2\dag}$
Zezheng Zhu,$^{3}$
James A. Hutchison,$^{4}$\\[0.35em]
Mia Wansbrough,$^{5}$
Nanfang Yu,$^{6\protect\#}$
Devi Stuart-Fox$^{7\protect\#}$\\[1em]
\normalsize
\begin{minipage}{0.92\textwidth}
\centering\footnotesize
$^{1}$School of Biosciences, University of Melbourne, Parkville, Victoria 3010, Australia\\[0.2em]
$^{2,3,6}$Department of Applied Physics and Applied Mathematics, Columbia University, New York, NY, USA\\[0.2em]
$^{4}$ARC Centre of Excellence in Exciton Science, School of Chemistry, University of Melbourne, Parkville, Victoria 3010, Australia\\[0.2em]
$^{5,7}$School of Biosciences, University of Melbourne, Parkville, Victoria 3010, Australia\\[0.6em]
$^{\dag}$These authors contributed equally to this work.\\
$^{\#}$Joint senior authors.\\
$^{\ast}$Corresponding author: laura.ospinarozo@unimelb.edu.au
\end{minipage}
\end{center}
\vspace{1.25em}

\begin{abstract}
Vivid colours in nature often arise from photonic nanostructures that have inspired diverse technologies. Yet most known examples fall within a limited set of structural themes. Here, we describe a biologically and optically novel structure in the bright green, violin-shaped stripes of the fiddler beetle \textit{Eupoecila australasiae}. The green colour is produced by a composite, hierarchical structure comprising dense arrays of microscopic, fin-like elements located beneath the cuticle. Each vertical fin, patterned with complementary lattices of nanospheres and indentations, can be approximated by two photonic crystal slabs mounted on a solid central core. Optical modelling shows that the fins are strongly iridescent, reflecting light with longer wavelengths near the normal and light with shorter wavelengths at oblique angles. However, disorder in fin orientation and filtering by the overlying cuticle converts the opaline cyan appearance of the fins into the bright diffuse green seen externally. Our work expands the known diversity of biological photonic nanostructures and offers new inspiration for biomimetic designs.
\end{abstract}

\section{Introduction}

Photonic nanostructures are widespread in nature and account for an astonishing diversity of optical effects that continue to inspire new technologies and applications. Many recurring photonic motifs found in nature are classified according to the dimensionality of their periodicity \cite{ref1}. For example, the multilayers of bright jewel beetles and the Bouligand (i.e. helicoidally twisted plywood) structures of metallic scarabs are classified as one-dimensional photonic crystals \cite{ref2,ref3,ref4,ref5,ref6,ref7}. Ordered rod lattices in iridescent bird feathers and ordered hole lattices in the spines of sea mice can be classified as two-dimensional photonic crystals \cite{ref8,ref9,ref10,ref11}. Opals and single gyroid structures found in butterflies and weevils are examples of three-dimensional photonic crystals \cite{ref12,ref13,ref14}. In these structures, colours are generated by architectures composed of periodically or aperiodically repeating units that extend across relatively large areas or volumes and can therefore be treated as effectively infinite systems with negligible boundary effects \cite{ref15}. However, the field of photonics has often explored compact architectures, in which finite dimensions are not merely a geometric limitation but are examined to realize enhanced control over light propagation and scattering \cite{ref15,ref16,ref17}. Some examples include waveguides, optical antennas, and photonic crystal slabs. Such architectures are widely studied in engineered photonic devices but are seldom considered in biological photonic structures. Their apparent rarity may arise because these architectures are difficult to develop, do not scale easily into conspicuous colour patches, or fail to generate robust visual signals. Equally possible is that they have simply escaped recognition. Either way, an intriguing question remains: can a compact photonic structure generate structural colour that remains conspicuous at the scale of an entire organism?

Among the most versatile compact photonic architectures are photonic crystal slabs. Structurally, a photonic crystal slab consists of a two-dimensional periodic arrangement of holes, pillars, or other nanostructures of finite height \cite{ref15,ref16}. When light interacts with a photonic crystal slab, three optical outcomes are possible. At certain wavelengths, the periodic structure prevents light from propagating through the slab, causing it to be reflected or back-scattered (photonic band gap). At other wavelengths, light remains confined within the slab and propagates along it as guided modes. Due to the finite thickness of the structure, some guided light becomes only partially confined and continuously radiates energy into the surrounding medium as leaky modes \cite{ref18}. This third behaviour is absent from infinitely thick two-dimensional photonic crystals, which support only photonic band gaps and guided modes. While leakage may not be desirable for waveguiding applications, it can be exploited to enhance outcoupling of light with specific colours \cite{ref19}. Several biological nanostructures resemble photonic crystal slabs morphologically, including the antireflective nipple arrays of moth eyes and the transparency-enhancing nanopillars in glasswing butterflies \cite{ref20,ref21}. However, these nanostructures have evolved deeply subwavelength dimensions to maximise light transmission rather than generate visible structural colour \cite{ref22,ref23}. To date, the only well-established biological examples of colour-generating photonic crystal slabs are the perforated silica frustules of diatoms \cite{ref24,ref25,ref26}. Yet diatoms are microscopic organisms, leaving open the question of whether similar optical principles can produce conspicuous structural colouration at the macroscopic scale.

Here, we show that the vivid green in the characteristic stripes of the Australian fiddler beetle, \textit{Eupoecila australasiae} (Scarabaeidae: Cetoniinae), is produced by arrays of densely packed, vertically oriented photonic crystal slabs on fin-like elements located beneath the translucent cuticle. By utilizing a hierarchical architecture, the system is capable of generating conspicuous colour patches at the scale of the whole organism. Remarkably, each hierarchical level introduces a distinct structural innovation and optical function: photonic crystal slabs mounted on micron-scale vertical fins produce highly iridescent optical effects; in addition, their assembly into a dense discontinuous array sums and spatially averages these responses into a cyan appearance; finally, an unusually translucent cuticle filters the reflected light, transforming it into the homogeneous green colouration observed in the characteristic fiddler stripes. This way of producing structural colour is unlike any previously described in nature. Its discovery is particularly surprising because the green markings of the fiddler beetle were assumed to be pigmentary owing to their rapid fading after death. Combining optical measurements, electron microscopy, and physical modelling, we reveal a complex form of structural colour that expands the repertoire of architectures available for bioinspired photonics.

\section{Results}

\subsection{Patterned photonic structures beneath the cuticle}

\begin{figure}[htbp]
\centering
\includegraphics[width=\textwidth,height=0.72\textheight,keepaspectratio]{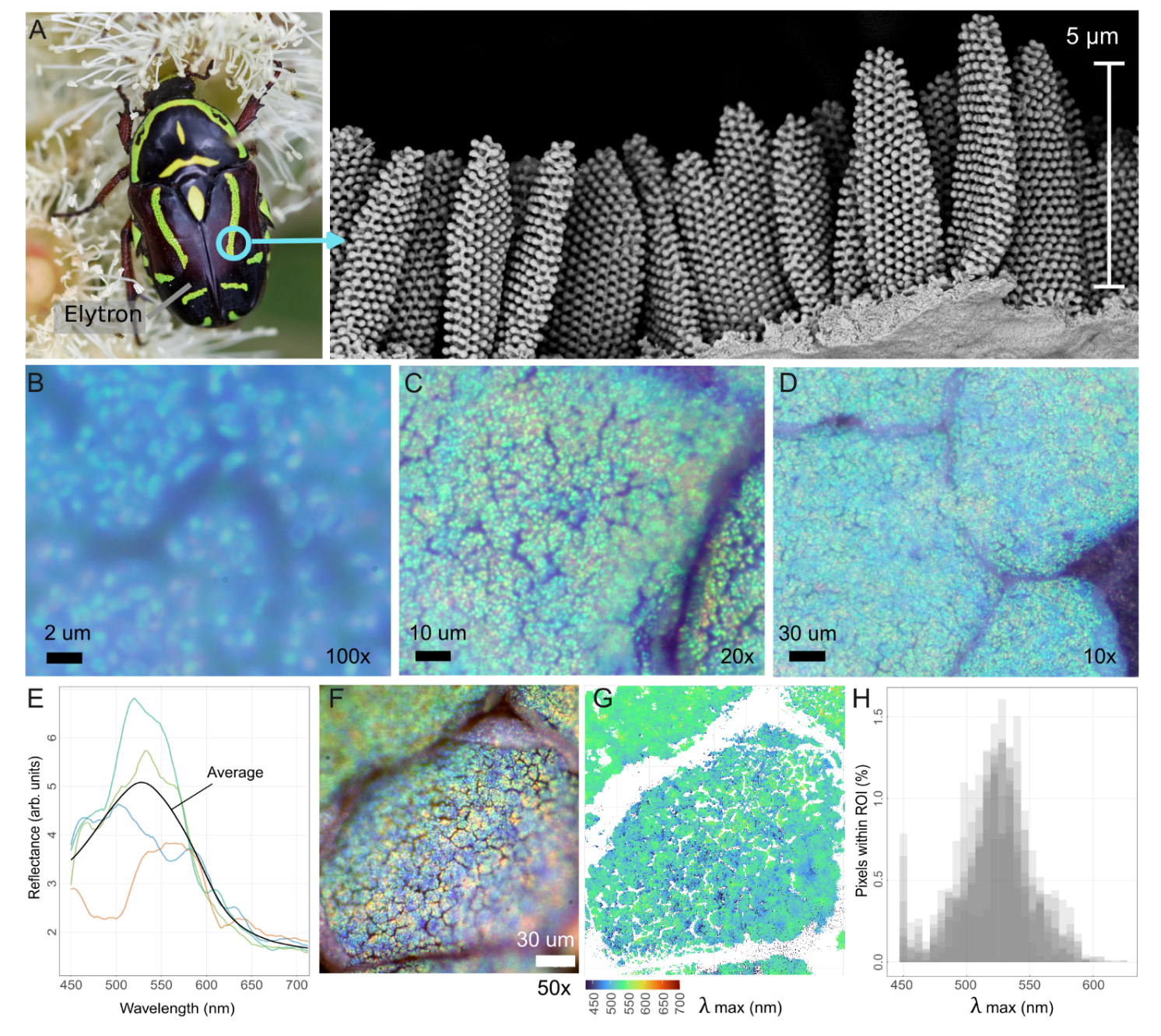}
\caption{\textbf{The green stripes in the fiddler beetle conceal microscopic fins with scale-dependent cyan reflectance.} \textbf{(A)} Left: The fiddler beetle, \textit{Eupoecila australasiae} Right: A scanning electron microscope (SEM) image showing the sideview of an array of vertical fins. \textbf{(B--D)} Bright-field microscopy images taken by 100x objective (NA=0.85, collection angle \textasciitilde{}60$^\circ$), 20x objective (NA=0.4, collection angle \textasciitilde{}23.5$^\circ$), and 10x objective (NA=0.25, collection angle \textasciitilde{}14.5$^\circ$), from left to right.  \textbf{(E--H)} Spectral measurements derived from hyperspectral imaging acquired with a 50$\times$ objective. \textbf{(E)} Representative reflectance spectra from individual pixels showing considerable variation in $\lambda$max and average spectrum from three hyperspectral images. \textbf{(F)} False-colour RGB image reconstructed from the hyperspectral cube, illustrating the characteristic pixel mosaic within a locally flat region where focus was optimized across wavelengths. \textbf{(G)} Spatial map of the wavelength of maximum reflectance ($\lambda$max) for the same region shown in F. Each pixel is coloured according to its $\lambda$max. Only pixels above a minimum reflectance threshold were included.  \textbf{(H)} The distribution of $\lambda$max across nine 31 $\times$ 31 pixel regions of interest sampled from three hyperspectral images}
\label{fig:main1}
\end{figure}

The green stripes of the fiddler beetle contain a dense array of discrete microscopic elements, hereafter termed fins (Fig.~\ref{fig:main1}A). The fins are absent from dark brown regions of the exoskeleton and their removal eliminates the green appearance of the stripes (Fig.~\ref{fig:s1}), indicating their role in colour production.  However, when viewed after dissection from the internal side of the exoskeleton, the fins themselves produce a shorter-wavelength opaline cyan appearance. Thus, the optical effect produced by the fins alone do not account for the homogeneous green appearance of the stripes.

The fins' appearance transformed markedly across spatial scales (Fig.~\ref{fig:main1}B-D). Specifically, the large numerical aperture of the 100x microscope objective (collection angle \textasciitilde{}60$^\circ$) captures oblique angle blue scattering, the 20x objective (collection angle \textasciitilde{}23.5$^\circ$) image captures the variable spectra of the opaline mosaic, while the 10x microscope objective (collection angle \textasciitilde{}14.5$^\circ$) shows the effect of summing and spatial averaging of colours produced by individual fins to form a more homogeneous cyan appearance. To approximate the reflectance produced by individual fins, we extracted spectra from single spatial pixels (1 pixel $\approx$ 247 $\times$ 247 nm) of hyperspectral data taken using a 50X objective (some examples in Fig.~\ref{fig:main1}E). Strong variation in the spectral shape and peak of neighbouring pixels suggests that nearby fins can generate distinct optical responses (Fig.~\ref{fig:main1}F-H). Spatial averaging over small areas was sufficient to produce a homogeneous, diffuse cyan appearance. In fact, integrated reflectance across approximately 100 pixels (corresponding to an area of \textasciitilde{}2.5 $\times$ 2.5 $\mu$m) was sufficient for the spectrum to converge to a broadband cyan spectrum, in all the studied regions of interest (Fig.~\ref{fig:s2}A). Consistent with this, back focal plane imaging of a small area 30 $\mu$m diameter spot revealed homogeneous scattering across a wide range of angles (Fig.~\ref{fig:s2}B). Together, these results show that the heterogeneous optical responses of individual fins rapidly converge to a homogeneous, diffuse macroscopic signal.

To capture the appearance of living specimens, all optical measurements described above had to be performed immediately after the beetles were sacrificed. This is because the colour in the fiddler stripes fades rapidly compared with most structural colours, which are renowned for their long-term stability and can persist for decades and even over geological timescales \cite{ref27,ref28,ref29}. Within approximately three days, the vivid green colour progressively faded despite the photonic fins remaining morphologically identifiable in electron microscopy images (Fig.~\ref{fig:s3}).

\begin{figure}[htbp]
\centering
\includegraphics[width=\textwidth,height=0.72\textheight,keepaspectratio]{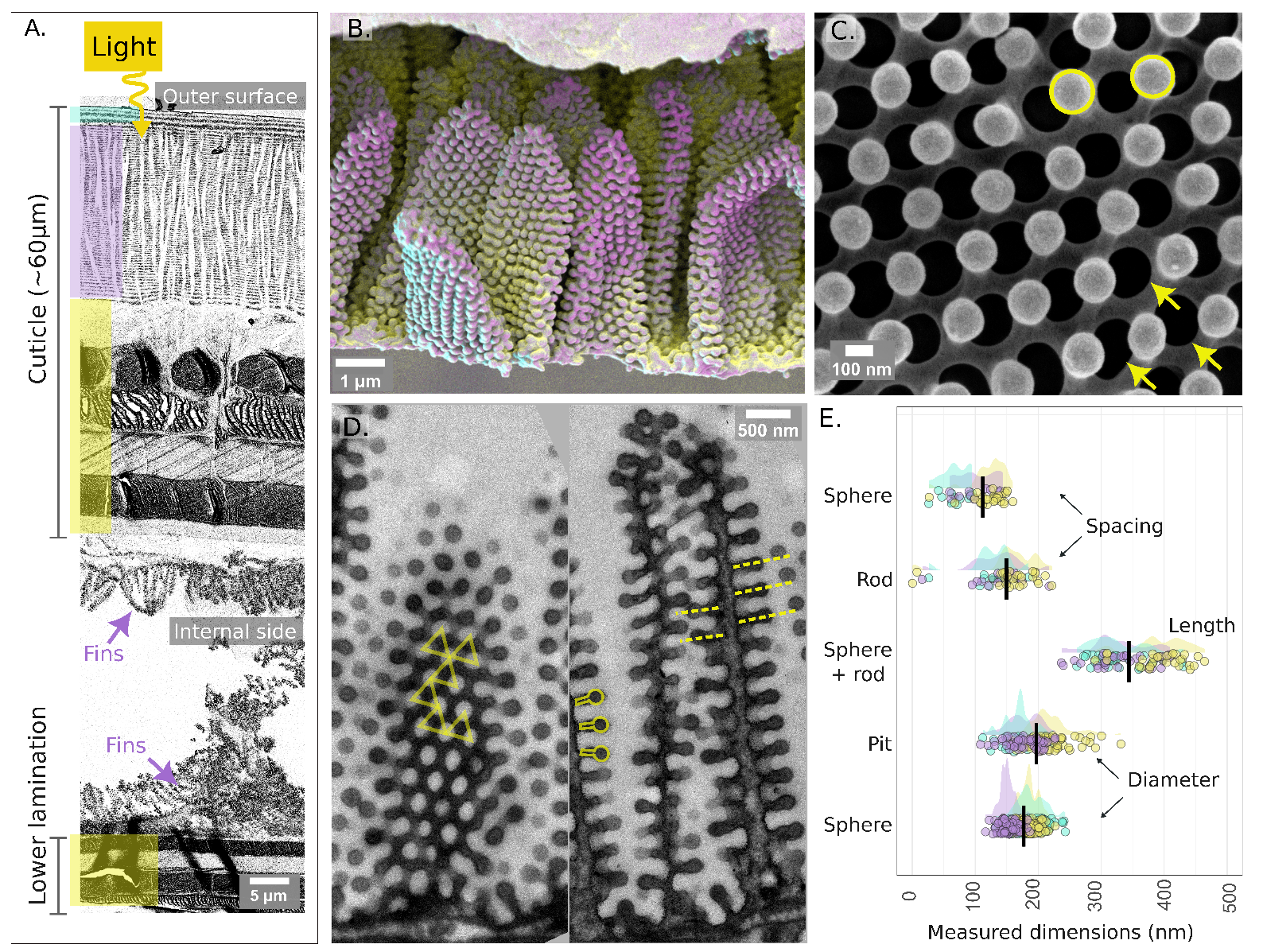}
\caption{\textbf{Morphological characterisation of the nanostructures.} \textbf{(A)} Contrast-inverted SEM image of a 200-nm-thick transverse section of the beetle elytra with shaded regions corresponding to the exocuticle (blue), mesocuticle (purple), and endocuticle (yellow). \textbf{(B)} SEM image showing a side view of a series of fins in different orientations. Artificially coloured by combining secondary and low-angle backscattered electron signals to distinguish surface and depth information. \textbf{(C)} SEM image of one face of a fin showing complementary triangular lattices of clavate spheres (circles) and indentations (arrows). \textbf{(D)} TEM micrographs of 90 nm-thick sections of the elytron showing cross-sectional views of fins in two different orientations. The triangular arrangement of the pitted lattice (triangles) is evident in parallel orientation and the sphere-on-rod geometry of the clavate lattice (circles) evident in perpendicular orientation with the lattices on opposite sides of a single fin are here offset by approximately half a period (dashed lines). \textbf{(E)} Dimensions of fin structural elements, including edge-to-edge spacing between adjacent spheres, edge-to-edge spacing between adjacent rods, sphere + rod length, pit diameter, and sphere diameter. Measurements were obtained from multiple regions across three EM images per beetle. Three beetles exhibiting slightly different stripe hues are shown in different colours.}
\label{fig:main2}
\end{figure}

Having identified the fins as the source of cyan reflectance, we next examined their location within the exoskeleton. The fins are intricately patterned units, approximately 5$\mu$m in height, lining the interior of specialized subsurface sacs located directly beneath the cuticle (Fig.~\ref{fig:main2}A). This configuration places the fins tens of microns below the external surface. In the elytra, these sacs are large and oval, extending between the dorsal and ventral cuticular layers, whereas in other body regions they are compressed (Fig.~\ref{fig:s4}). As a result, opposing fin-lined surfaces can be widely separated (\textasciitilde{}40$\mu$m) or in near contact.

Detailed characterisation of the fin ultrastructure revealed that they are broadly upright and perpendicular to the local body surface but vary subtly in curvature and orientation (Fig.~\ref{fig:main2}B). Each fin consists of a flattened, tapered element whose opposing faces contain complementary triangular lattices (Fig.~\ref{fig:main2}C). On each side, a clavate lattice composed of nanospheres supported by rods is paired with an offset pitted lattice of nanoindentations within the central core. The lattices on opposite faces share the same periodic arrangement but can be offset from one another (Fig.~\ref{fig:main2}D). All spacing and characteristic dimensions of the lattice features are below 500 nm, placing them within the submicron scale required for interactions with visible light (Fig.~\ref{fig:main2}E).

\subsection{Light-matter interactions in the fins}

From an optics perspective, each fin consists of two photonic crystal slabs mounted on either side of a solid central core (Fig.~\ref{fig:main2}B). Each photonic crystal slab comprises two complementary triangular lattices formed by clavate protrusions and pitted indentations (Fig.~\ref{fig:main2}D). Because the fins are oriented roughly perpendicular to the outer cuticle surface (Fig.~\ref{fig:main2}A-B), light couples into the fins from their ends largely in-plane with respect to the photonic crystal slabs and is back-scattered following wavelength-dependent pathways: light with long wavelengths that falls within the bandgaps of the photonic crystal slab is reflected at small angles with respect to the surface normal, and light with short wavelengths (i.e., green and blue) is back-scattered into leaky waves at large angles. This result strongly aligns with the spectral measurements obtained experimentally (Fig.~\ref{fig:main1}C), in which broader angular collection enhanced shorter-wavelength contributions to the reflectance spectrum of the fin array.

\begin{figure}[htbp]
\centering
\includegraphics[width=\textwidth,height=0.72\textheight,keepaspectratio]{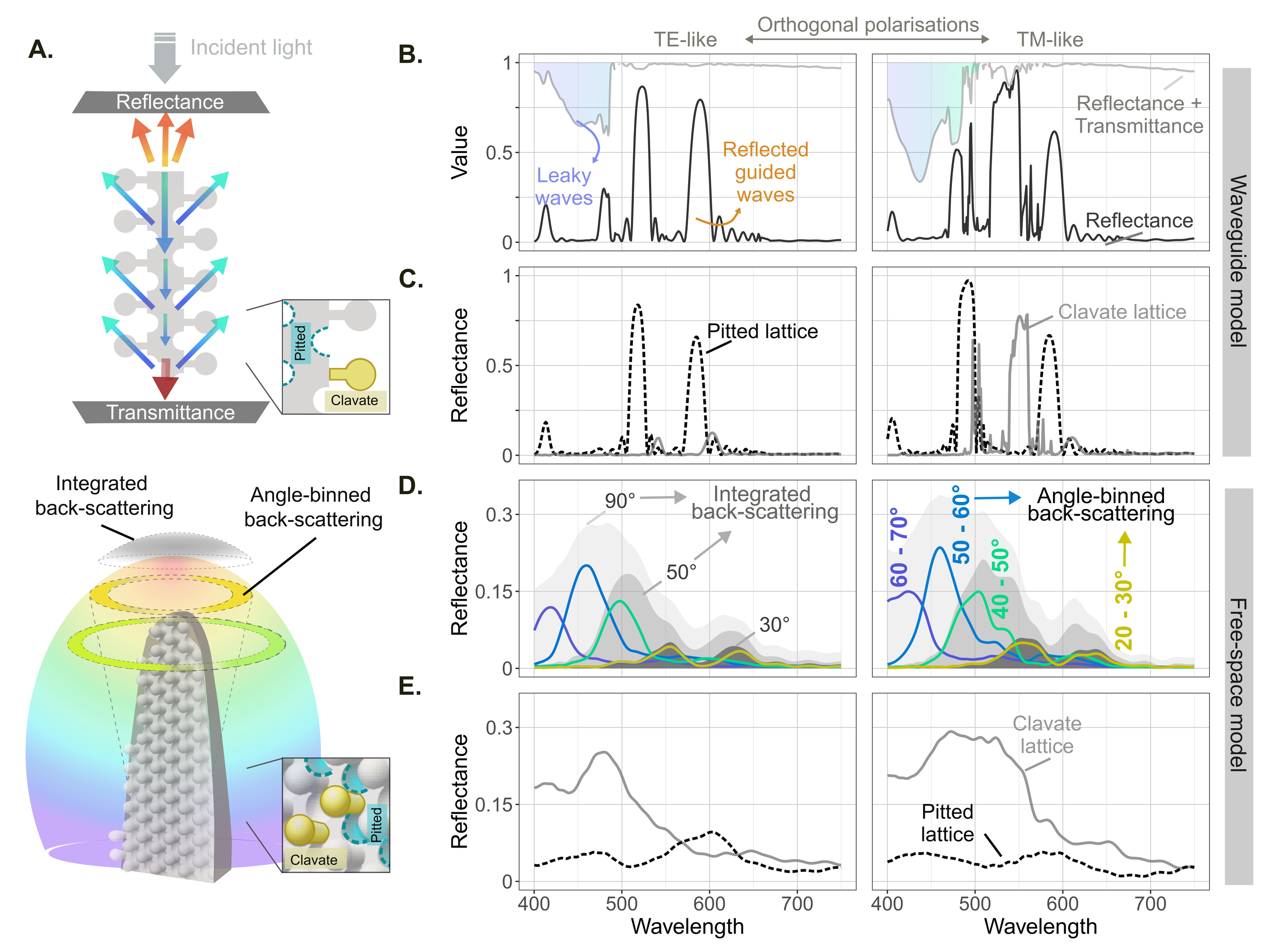}
\caption{\textbf{Simulations of the interaction between light and the fin.} \textbf{(A)} Schematic of the two modelling approaches. In both, the structure comprises a chitin core overlain on its two surfaces by pitted (air hemispheres) and clavate (chitin spheres and cylinders) triangular lattices with a lattice constant \textasciitilde{}280 nm (insets). Upper: waveguide model used to monitor guided reflectance and transmission as well as leaky waves. The waveguide extends infinitely into and out of the page. Lower: model used to calculate integrated and angle-binned backscattering of a single fin in free-space. Coloured arrows and background qualitatively illustrate the wavelength-dependent scattering. \textbf{(B)} Results of the waveguide model using lossless chitin (n=1.55, k=0). Black curves show spectra corresponding to reflected guided waves propagating along the core. Light that is neither reflected nor transmitted (i.e., one minus the sum of reflectance and transmittance) represents optical power coupled into leaky waves. \textbf{(C)} Reflectance spectra of the waveguide model showing distinct contributions from each of the two lattices. \textbf{(D)} Results of the free-space model showing angle-binned back-scattered optical power displayed in 10$^\circ$ intervals and represented in human- perceived colour. The shaded region represents integrated backscattering over the upper hemisphere. \textbf{(E)} Results of the free- space model showing integrated back-scattered optical power due to the pitted and clavated lattices. Integrated optical power in \textbf{(D)} and \textbf{(E)} was normalized as described in Methods.}
\label{fig:main3}
\end{figure}

We investigated the optical properties of the fins with a waveguide model and a free-space model, both using full-wave simulations. In the waveguide model, we studied an infinitely wide fin excited by a slab waveguide mode (Fig.~\ref{fig:main3}A). This approach enabled us to clearly distinguish between in-plane reflected light (i.e., light waves that fall within the photonic band gaps), out-of-plane back-scattered light (i.e., leaky waves), and forward-propagating guided waves. In the free-space model, a structure closely resembling a fin was created and excited by a plane wave (Fig.~\ref{fig:main3}A). This approach enabled us to gain a comprehensive understanding of the iridescent properties (i.e., angle-dependent back-scattering) of the fin. In both cases, the surface features of the fin were modelled by chitin cylinders, chitin spheres, and air hemispheres in a perfect triangular lattice, using the mean dimensions obtained from electron microscopy measurements (Fig.~\ref{fig:main2}E and Table~\ref{tab:s1}). For both models, results were calculated for two orthogonal polarizations of injected light: one in which the electric field is oriented parallel to the flat faces of the fin (i.e., transverse-electric or TE excitation), and the other in which the electric field is oriented perpendicular to the flat faces of the fin (i.e., transverse-magnetic or TM excitation). While the free-space model is more realistic and allows for direct comparison with measurements, the more abstract waveguide model is useful to understand the underlying optical mechanisms. The results of the two models are complementary; for example, light waves at short wavelengths that leak out of the core in the waveguide model correspond to those scattered at large angles in the free-space model (Fig.~\ref{fig:main3}B and D).	Fig.~\ref{fig:main3}. Simulations of the interaction between light and the fin. (A) Schematic of the two modelling approaches. In both, the structure comprises a chitin core overlain on its two surfaces by pitted (air hemispheres) and clavate (chitin spheres and cylinders) triangular lattices with a lattice constant \textasciitilde{}280 nm (insets). Upper: waveguide model used to monitor guided reflectance and transmission as well as leaky waves. The waveguide extends infinitely into and out of the page. Lower: model used to calculate integrated and angle-binned backscattering of a single fin in free-space. Coloured arrows and background qualitatively illustrate the wavelength-dependent scattering. (B) Results of the waveguide model using lossless chitin (n=1.55, k=0). Black curves show spectra corresponding to reflected guided waves propagating along the core. Light that is neither reflected nor transmitted (i.e., one minus the sum of reflectance and transmittance) represents optical power coupled into leaky waves. (C) Reflectance spectra of the waveguide model showing distinct contributions from each of the two lattices. (D) Results of the free-space model showing angle-binned back-scattered optical power displayed in 10$^\circ$ intervals and represented in human-perceived colour. The shaded region represents integrated backscattering over the upper hemisphere. (E) Results of the free-space model showing integrated back-scattered optical power due to the pitted and clavated lattices. Integrated optical power in (D) and (E) was normalized as described in Methods.

We first modelled a fin as a waveguide, infinite in width, \textasciitilde{}4.5 $\mu$m in length, and composed of a solid core \textasciitilde{}360 nm in thickness with each of its two surfaces patterned with a set of clavate and pitted lattices (emulating the morphology in Fig.~\ref{fig:main2}C). Based on our structural studies (e.g., Fig.~\ref{fig:main2}D), we offset the patterns on the two surfaces to form an AB stacking (i.e., points of the triangular clavate lattice on one surface are aligned with triangular interstices of the clavate lattice on the other surface). The waveguide was illuminated with either the fundamental TE or TM slab waveguide mode of the solid core and reflected (R) and transmitted (T) guided waves were monitored. Assuming lossless (non-absorbing) chitin, the simulation results indicated that at wavelengths greater than \textasciitilde{}480 nm light is largely confined within the waveguide (R+T $\cong$ 1; Fig.~\ref{fig:main3}B); specifically, the waveguide strongly reflects light within three bands between 480 and 600 nm, transmitting the remainder.  Conversely, light with wavelengths below \textasciitilde{}480 nm is not fully reflected nor transmitted (R+T<1; Fig.~\ref{fig:main3}B). Since our materials model did not consider absorption, the missing fraction of guided modes corresponds to leaky waves scattered into free space by the surface features. A materials model accounting for chitin absorption (using reported complex optical refractive indices (30)) resulted in increased attenuation at shorter wavelengths and similar optical scattering outside of the core within the three bands (Fig.~\ref{fig:s5}). We found that reflection is stronger for TM than for TE guided modes, due to the anisotropic geometry of the waveguide. Using further reduced models in which the clavate and pitted lattices were simulated separately, we found that they each contribute to distinct reflection bands: the pitted lattice is essential for reflection in both TE and TM polarisations, whereas the clavate lattice is more relevant in reflecting the TM mode (Fig.~\ref{fig:main3}C and Fig.~\ref{fig:s5}).

\begin{figure}[htbp]
\centering
\includegraphics[width=\textwidth,height=0.72\textheight,keepaspectratio]{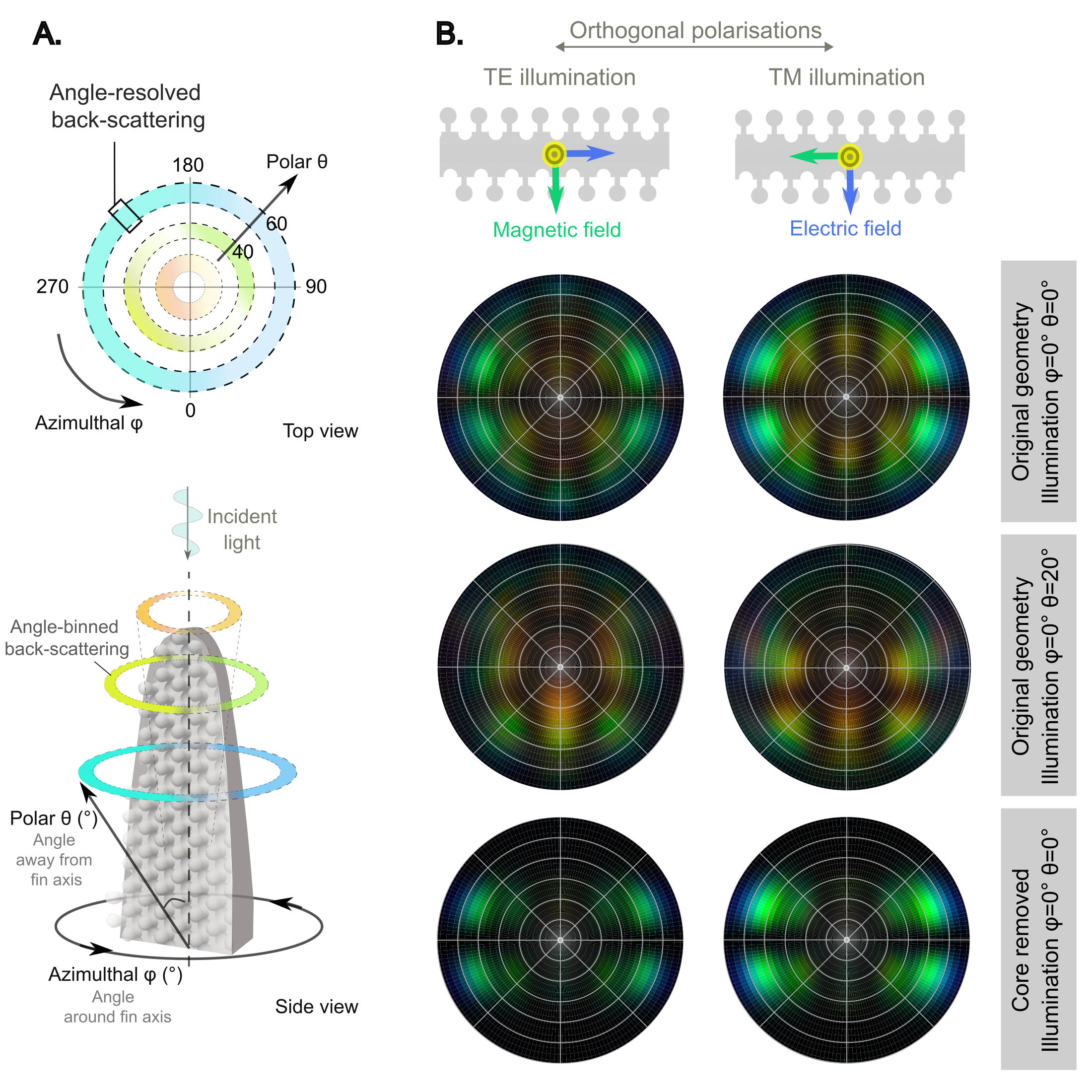}
\caption{\textbf{The architecture of the fin redistributes light waves into distinct angular pathways depending on polarisation and illumination angle.} \textbf{(A)} Schematic representation of the angular coordinates used in the free-space model. Here, back-scattering is resolved for every combination of polar (3$^\circ$ intervals) and azimuthal (1$^\circ$ intervals) angles relative to the fin axis using near-to- far-field projection. Results are displayed as top views centred on the fin axis and truncated at a polar angle of 60$^\circ$, where the dominant scattering features are observed. Colours represent the human-perceived colour derived from the simulated spectra. \textbf{(B)} Simulated angular distributions of back-scattered light from a single fin for two orthogonal incident polarization states (TE and TM). Schematics above each column indicate the orientation of the incident electric and magnetic field components relative to the fin geometry.}
\label{fig:main4}
\end{figure}

We next used full-wave simulations to study how free-space injected light is scattered by an individual fin. The fin was modelled as a \textasciitilde{}360-nm-thick half-elliptical-cylinder where each of its two flat surfaces is patterned with a set of clavate and pitted lattices. This fin model was erected vertically and illuminated by a plane-wave source from above, and we calculated integrated and angle-resolved back-scattering using near-field-to-far-field projection. The results showed that the spectral peaks shift to shorter wavelengths as the scattering angle increases (Fig.~\ref{fig:main3}D), consistent with the experimental results (Fig.~\ref{fig:main1}D-E). Simulations using either the pitted or the clavate lattice alone suggested that the latter plays a major role in light scattering (Fig.~\ref{fig:main3}E). Simulated scatterometry images \cite{ref31,ref32} are viewable as angle-resolved polar plots of back-scattering represented as human-perceived colours (Fig.~\ref{fig:main4}). Specifically, light at longer wavelengths is back-scattered weakly at smaller angles from the fin axis; light at shorter wavelengths is more strongly scattered across a broader angular range via leaky-wave pathways, resulting in the dominant cyan appearance of the fin array. Furthermore, the angular distribution of back-scattered light changed between incident polarization states, consistent with the anisotropic geometry of the fin. Altering the angle of incidence subtly redistributed the scattered light.Fig.~\ref{fig:main4}. The architecture of the fin redistributes light waves into distinct angular pathways depending on polarisation and illumination angle. (A) Schematic representation of the angular coordinates used in the free-space model. Here, back-scattering is resolved for every combination of polar (3$^\circ$ intervals) and azimuthal (1$^\circ$ intervals) angles relative to the fin axis using near-to-far-field projection. Results are displayed as top views centred on the fin axis and truncated at a polar angle of 60$^\circ$, where the dominant scattering features are observed. Colours represent the human-perceived colour derived from the simulated spectra. (B) Simulated angular distributions of back-scattered light from a single fin for two orthogonal incident polarization states (TE and TM). Schematics above each column indicate the orientation of the incident electric and magnetic field components relative to the fin geometry.

The characteristic spectral backscattering fingerprint of a single fin emerges from the collective contribution of multiple features. Sensitivity analyses identified the triangular lattice constant as the parameter exerting the strongest influence on the spectral peak location (Fig.~\ref{fig:s6}). The spectrum is overall resilient to changes in the feature dimensions, such as sphere and pit diameters and cylinder length, but sensitive to the complete removal of any of these two lattices (Fig.~\ref{fig:main3}). The core of the fin is not only the physical foundation supporting the clavate and pitted lattices, but also a critical optical component. Its thickness (\textasciitilde{}360 nm) supports confined guided waves that mediate the interaction between light and the photonic crystal slabs on its two surfaces. This role is demonstrated by simulations in which removing the core alters the back-scattering pattern and eliminates long-wavelength reflectance near the fin axis (Fig.~\ref{fig:main4}B). The central core also contributes to the stability of the back-scattering spectra across illumination angles up to \textasciitilde{}50$^\circ$ (Fig.~\ref{fig:s7}). The tapered shape of the fin likely enhances the coupling of light into and out of the structure and thus the efficiency of back-scattering, analogously to enhanced coupling in tapered waveguides \cite{ref33}. Our simulations showed that increasing fin height increased the total intensity of back-scattered light, as taller fins contain more periods of the photonic crystal slab. However, these gains progressively diminished, approaching a plateau. The range of fin heights observed in the microscopy images (4.2-5.2 $\mu$m; Table~\ref{tab:s1}), lies close to this plateau, where near-maximal back-scattering is achieved with minimal additional material (Fig.~\ref{fig:s8}). Finally, beyond the morphological parameters of the fin, refractive-index contrast also plays a critical role. Since the optical effect relies on the contrast in refractive index between the fin and its surroundings, postmortem changes affecting either component could explain the loss of colour despite the persistence of the nanostructure (Fig.~\ref{fig:s9}).

The organisation of multiple fins within an array introduces additional structural variables that influence optical appearance. Small variations in fin orientation from the vertical direction may alter the dominant light-collection pathways between in-plane back-scattering at long wavelengths and out-of-plane back-scattering at short wavelengths, explaining the blue-to-yellow heterogeneity observed in the fin array (Fig.~\ref{fig:main1}C). Interestingly, the free-space model of two adjacent fins showed that the secondary scattering, i.e., light scattered by one fin entering and being subsequently scattered by the other fin back to free-space, is negligible (Fig.~\ref{fig:s10}). In contrast, a double layer of opposing fin-lined surfaces, as present in the fiddler beetle elytron (Fig.~\ref{fig:main2}A), results in an increase in back-scattering, but not a significant change in spectral location (Fig.~\ref{fig:s11}). Overall, the collective effect of fins in an array is a spatial and angular mixture of the iridescent contributions from individual fins, which results in a cyan appearance, consistent with observations from the inner side of the cuticle but not with the green appearance observed externally.

\subsection{The cuticle as a colour filter}

\begin{figure}[htbp]
\centering
\includegraphics[width=\textwidth,height=0.58\textheight,keepaspectratio]{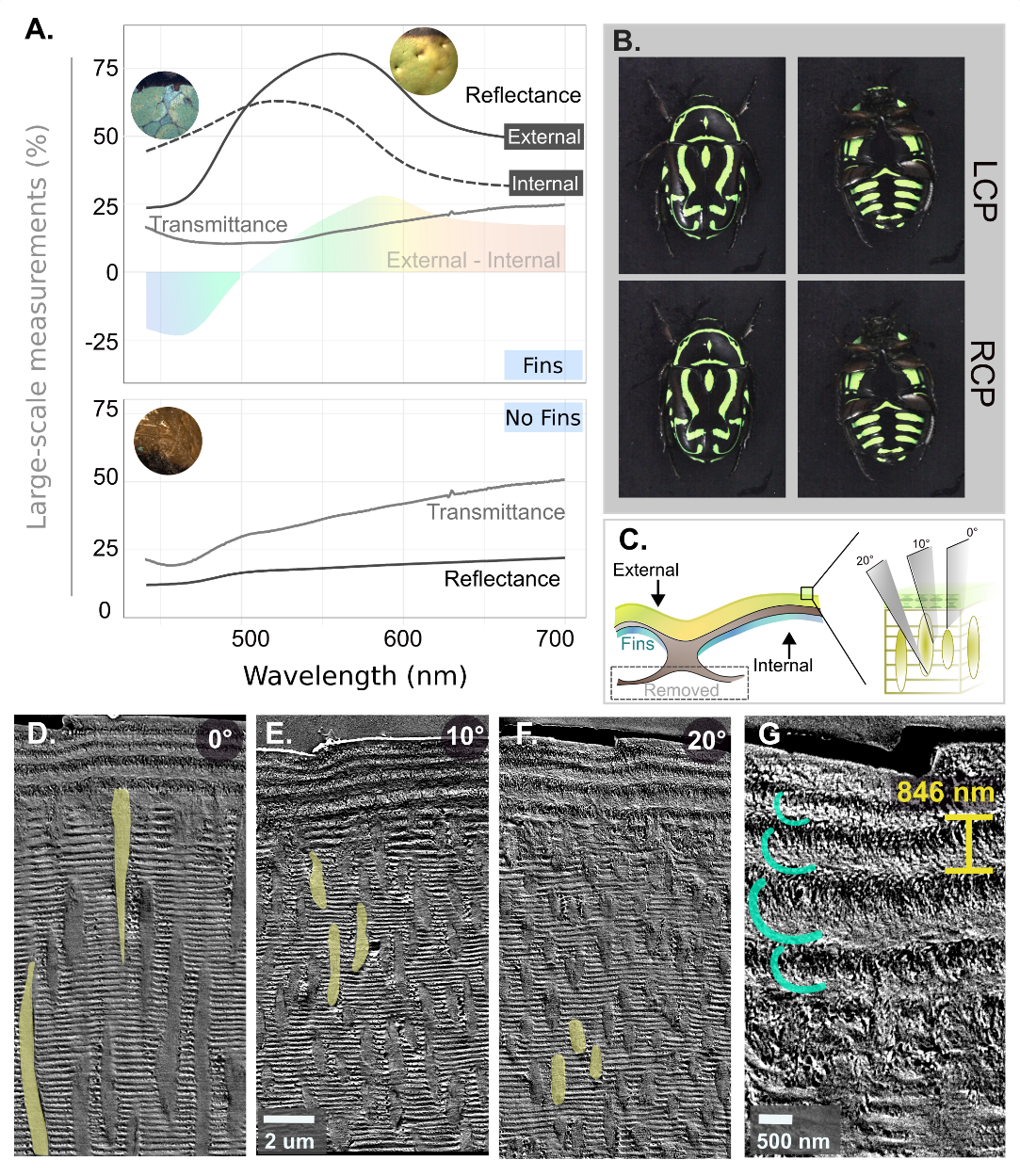}
\caption{\textbf{Optical properties and structural organisation of the fiddler beetle cuticle.} \textbf{(A)} Reflectance and transmittance spectra measured from a green stripe of the elytron (mean of two individuals). Top: Reflectance (black) and transmittance (grey) measured from the external side of the intact elytron and internal side after removal of the lower lamination (measurement schematic shown in \textbf{(C)}. The shaded region (external -- internal reflectance) represents the wavelength-dependent spectral modification introduced by the overlying cuticle. Bottom: Reflectance and transmittance of the cuticle alone, obtained after manual removal of the fin array. \textbf{(B)} Photographs of \textit{E.\ australasiae} under left- and right-handed circularly polarised light, showing no polarisation-dependent reflectance. \textbf{(C)} Schematics of the experimental approaches used to isolate the optical contributions of the cuticle and fin array (left) and reconstruct the three-dimensional architecture of the fin array from oblique SEM sections at 0$^\circ$, 10$^\circ$, and 20$^\circ$ from the normal (right). \textbf{(D--F)} TEM sections obtained at oblique angles, showing changes in pore shape and density due to sectioning geometry (scale bar: 2 $\mu$m for all three). \textbf{(G)} Higher magnification TEM image of the outer cuticle layers, showing a thin region packed with chiral fibres (semicircles) of apparent large pitch values.}
\label{fig:main5}
\end{figure}

The discrepancy between the opaline cyan appearance of the fin array and the homogeneous green appearance of the stripes is caused by the cuticle that overlays the fin array (Fig.~\ref{fig:main5}A). To isolate the optical contribution of the cuticle, we measured the reflectance of the intact elytron, capturing the combined optical response of the fin array and the overlying cuticle. This produced a broad green reflectance peak centred at $\lambda$ = 550 nm. After removal of the lower lamination, the fin array was exposed from the internal side, allowing its reflectance to be measured in isolation (Fig.~\ref{fig:main5}C). The reflectance of the fin array peaked at a shorter wavelength $\lambda$ = 530 nm, closely matching the average microscale reflectance measured from individual fins (Fig.~\ref{fig:main1}E). Consequently, the difference between the external and internal reflectance spectra isolates the wavelength-dependent contribution of the cuticle (Fig.~\ref{fig:main5}A- shaded area). This difference reveals an approximately 25\% reduction in reflectance at wavelengths below 500 nm when the overlying cuticle is present. Remarkably, the transmittance spectrum of the isolated cuticle exhibits a nearly identical attenuation below 500 nm (Fig.~\ref{fig:main5}A- bottom panel), demonstrating that the cuticle acts as a spectral filter by selectively absorbing short wavelengths as light passes through it before and after reflection from the fin array. Overall, these results indicate that the cuticle acts as a colour filter, shifting the opaline cyan reflectance of the fins to the observed green hue of the beetles' markings.

The cuticle of the fiddler beetle differs markedly from the optically active cuticles typical of scarabs. It shows little overall reflectance, and it does not preferentially reflect one handedness of circular polarisation over the other, at least in the visible range (Fig.~\ref{fig:main5}B). Rather than contributing substantially to the reflected signal, the cuticle appears to be specialised for high optical transmission, despite being approximately 60 $\mu$m thick (Fig.~\ref{fig:main2}A). This unusual optical role suggests that the fiddler beetle cuticle is different from the optically active cuticles typical of many scarabs \cite{ref34,ref35,ref36}.

To determine if chiral fibres were present in the fiddler cuticle, we performed oblique sectioning of the cuticle at defined angles relative to the surface normal (Fig.~\ref{fig:main5}C-F). We found that only the outermost layers of the cuticle display evidence of chiral fibre organisation, and even there, the pitch is substantially larger than the maximum pitch expected to reflect visible light waves (Fig.~\ref{fig:main5}G). Instead, the cuticle is dominated by a porous, non-chiral architecture, likely corresponding to the mesocuticle, a transitional zone between the typically colour-associated exocuticle and the mechanically resistant endocuticle (Fig.~\ref{fig:main5}D-F). Thus, although chiral architectures are present in the fiddler beetle cuticle, they do not contribute to colour production. Overall, the unusual structural organisation of the cuticle appears to minimise its reflective contribution, allowing light to penetrate to the underlying fins.

\section{Discussion}

The deceptively simple green stripes of the fiddler beetle emerge from an elaborate form of structural colour. At its core is a previously undescribed biophotonic architecture comprising two opposing photonic crystal slabs within a single fin-like microstructure. This is the first biological example of photonic crystal slabs generating colour at the macroscopic scale of an animal. At the micron scale, opposing photonic crystal slabs generate angle-dependent optical responses through photonic band gaps and leaky-wave scattering. At the mesoscale, these slabs are organised into dense discontinuous arrays, a departure from the continuous optical media characteristic of most biological structural colours \cite{ref37,ref38}. Here, individual fins act as iridescent colour pixels, while their collective effects and disorder spatially sum and average to create a homogeneous cyan appearance. Finally, a cuticle that remains unusually translucent despite its thickness acts as a spectral filter, transforming the cyan mosaic into the homogeneous green of the fiddler stripes. Remarkably, such an intricate structural system produces ephemeral colour. We suggest that this apparent contradiction arises from the exceptional sensitivity of this photonic system to changes in refractive-index contrast, as shown by our simulations. The unusual photonic structures of the fiddler beetle suggest that unconventional routes to colour production may be more widespread in nature than currently recognised.

A single fin is already a sophisticated optical device because it integrates multiple independently tuneable structural features. Sphere-on-rod lattices, pitted lattices and complementary lattices all occur independently in other biological photonic systems \cite{ref39,ref40,ref41}, although in different structural contexts. Their integration within a single fin introduces numerous structural parameters that influence the reflected spectrum, albeit to different extents. For example, the spectral signature of an individual fin proved relatively robust to variation in the size of the lattice elements (pits, spheres and rods) and the orientation of the triangular lattice, yet highly sensitive to lattice spacing. Although this pattern is expected for engineered photonic crystal slabs, it highlights a simple design strategy for biomimetic analogues, where lattice spacing could be used to generate a library of colours while preserving the underlying architecture. Another notable finding is that the observed fin lengths lie close to the point of diminishing optical returns. Although this observation provides a practical guideline for biomimetic design, it remains unclear whether this apparent limit reflects selection for optical performance, anatomical constraints, or other evolutionary pressures. The influences of lattice geometry, lattice disorder, tapering of the fin, core dimensions, disorder in fin orientation, and fin bending remain unexplored. Understanding how these features interact to shape angle- and polarisation-dependent optical responses may provide new opportunities for bioinspired photonics.

The discontinuous fin array of the fiddler beetle enables structural disorder and spatial averaging. The transformation of highly directional optical effects into diffuse visual appearances is not uncommon in nature \cite{ref42,ref43}. For example, pointillistic mixing contributes to camouflage against sandy backgrounds in tiger beetles \cite{ref1,ref44}, while other forms of structural disorder facilitate camouflage against foliage in gold dust weevils \cite{ref45}, diamond weevils \cite{ref46} and lycaenid butterflies \cite{ref47}. In these cases, a consistent appearance from any viewing angle is ecologically advantageous. In fiddler beetles, the green patterning could enhance camouflage by breaking up the body outline (disruptive camouflage), but the functional significance of their diffuse reflectance remains speculative. From a photonics perspective, disordered photonic crystal slabs provide the opportunity to create stable spectral profiles across viewing angles, a sought-after property of biomimetic structural colour. A more detailed characterization of disorder within the fin array---including variations in fin bending, orientation, spacing, and alignment---could help disentangle which structural features contribute most strongly to the observed appearance. Such information may help identify design principles useful for engineered photonic systems.

The organisation of leaky-mode resonators in the fiddler beetle provides a biological proof of concept for an alternative photonic architecture, revealing interesting design principles for engineered photonic devices. Guided-mode resonances are increasingly used in dielectric metasurfaces, guided-mode filters and structural colour devices \cite{ref48,ref49,ref50}. Fragmenting a photonic crystal slab into discrete resonators allows individual elements to be independently tuned for specific spectral, angular or polarisation effects. Local programmability is already exploited in pixelated photonic devices \cite{ref51,ref52,ref53}. This modularity, however, requires efficient coupling light into each slab. Orienting the slabs vertically, as observed in the fiddler beetle, aligns the guiding layer with normally incident light, allowing all resonators to be excited simultaneously. Although such architectures are not readily compatible with conventional planar nanofabrication, emerging three-dimensional fabrication and post-fabrication reconfiguration approaches, including membrane transfer and self-folding \cite{ref49,ref54}, may provide practical implementation routes. Importantly, our modelling showed that not all aspects of the biological morphology contribute equally to the optical response, so preserving the underlying optical principles may be more important than reproducing the exact biological morphology. This opens opportunities to develop simplified, fabrication-compatible geometries that exploit the advantages of vertically oriented discontinuous photonic crystal slab arrays

The cuticle of the fiddler beetle appears to represent an evolutionary repurposing of a typical scarab reflector. It is generally assumed that scarab beetles lacking circularly polarised reflectors have lost the underlying chiral helicoidal architecture \cite{ref5}, yet the fiddler beetle retains a reduced, optically inactive helicoid, potentially representing an evolutionary remnant from the ancestor of flower chafers (subfamily Cetoniinae). The fiddler cuticle is dominated by pore canals whose optical function remains uncertain. Although their roles in cuticle development, mechanical support and lipid transport are well established \cite{ref55,ref56,ref57,ref58}, conflicting evidence suggests that they may either increase scattering or enhance transparency \cite{ref56,ref59,ref60}. The fiddler cuticle also selectively absorbs shorter wavelengths. Whether this filtering results from a single pigment or from the combined absorption of multiple molecules within the chitin--protein matrix remains unknown \cite{ref61,ref62,ref63,ref64}. While transmissive cuticles have been reported in other scarabs \cite{ref65}, their integration with colour-producing nanostructures confined within specialised subsurface sacs is striking. This organisation may facilitate the evolution of sharply defined colour patches without requiring abrupt structural transitions within the cuticle. Overall, the highly modified cuticle of the fiddler beetle offers an excellent opportunity for comparative studies aiming to understand the evolution of biophotonic systems.

The discovery of a remarkable photonic architecture in the fiddler beetle stripes gives rise to a rich set of questions spanning optics, development, evolution, and bioinspired applications. For instance, the function, evolution and development of coloration in fiddler beetles remains unknown. Although the close agreement between simulations and experiments provides robust support for the proposed optical mechanism, further work is needed to characterise the material composition of the fins \cite{ref61,ref66} and determine how physiological changes may influence their optical performance \cite{ref67,ref68}. Comparative studies across cetoniines \cite{ref69} may help determine the prevalence of this architecture among related species and selective pressures that have shaped its evolution. Likewise, exploring whether the hierarchical integration of multiple photonic mechanisms can be reproduced in synthetic materials may reveal new opportunities for scalable optical technologies that exploit controlled disorder and hierarchical optical control. Overall, the discovery of a new photonic mechanism in the fiddler beetle highlights the underexplored diversity of biological light-manipulating systems and their potential for bioinspired photonics.

\section{Materials and Methods}

Electron microscopy, optical measurements, and computational modelling were combined to characterise the structure and optical function of the fins. Morphological measurements extracted from electron microscopy images were used to parameterise optical models, which were then compared with experimental reflectance and transmittance measurements. Additional analyses of the overlying cuticle were performed to evaluate its contribution to colour production.

\subsection{Specimens and Sample Preparation}

Fiddler beetles were collected on roadsides near Yarraman State Forest, Queensland, Australia and kept under laboratory conditions at 27$^\circ$C and 12:12 hour light: dark cycle. Elytra from recently deceased beetles were dissected for optical measurements. Fragments were subsequently isolated using a 1 mm tissue punch for microscopy. Whenever light microscopy was used to measure optical properties, 2 mm diameter fragments of the elytron were mounted on microscope slides and secured with a small drop of transparent nail polish applied to one edge only, to avoid altering optical characteristics. For SEM imaging, similarly sized fragments were mounted directly onto SEM stubs. For TEM analysis, 2 mm fragments were placed in a fixative solution of 2.5\% glutaraldehyde in phosphate-buffered saline (PBS).

To isolate the contribution of the fins, green and brown regions of the elytra were isolated and the lower lamination was carefully removed from the ventral side using a fine blade and tweezers to expose the underlying fins where present. In half of the samples, the fin layer was then removed using the smallest and softest tip of a jeweller's rotary tool, taking care not to damage any other structures. These steps enabled comparison between intact and altered regions, and between the two sides of the exoskeleton in the same elytron.

\subsection{Optical Characterisation}

The reflectance and transmittance of different regions of the beetle elytra were measured using a combination of spectroscopic techniques.

Macroscopic reflectance was measured at normal incidence using a bifurcated fiber optic cable to two spectrometers (AvaSpecNXS2048CL, 300-1100 nm; AvaSpec-MINI-NIR256, 975-1700 nm). The illumination source was a 10 W halogen light source (AvaLight-Hal-S-Mini2). Each region was measured five times per specimen. Measurements were recorded with Avantes software AvaSoft (version 8), and calibrated against a 99\% diffuse reflectance Spectralon standard (Labsphere, North Sutton, NH, USA). Spectral data were processed using the R package pavo \cite{ref70}.

Microscopic reflectance was measured using a hyperspectral imaging system [Photon etc IMA and Andor Ixon 888 EMCCD] coupled to a microscope [Nikon LV100] with 50$\times$ and 100$\times$ air objectives [Nikon BD]. Spectra were calibrated against a 40\% dark mirror standard, and the resulting spectral cubes were processed using PHySpec software (V2.28 -- Photon etc. Inc. Montreal Quebec Canada). Angle-resolved reflectance at the microscopic scale was measured with microscatterometry (back focal plane imaging -- Spectrometer Acton SP2300 -- Princeton Instruments) over an area of \textasciitilde{}20 $\mu$m, within the visible spectrum (limited by the wavelength sensitivity of silicon-based detectors). The setup included a highly focused broadband tungsten-halogen light source (Ocean Optics HL-2000) coupled to a microscope objective [Olympus air objective, 50X, 0.90]. The beam was collimated and expanded to fill the objective's back aperture, allowing illumination at angles from 0$^\circ$ to 69$^\circ$ from the surface normal. Scattered light was collected through the same objective, and an image of the back focal plane was projected onto a CCD camera (PIXIS 1024B, Princeton Instruments) via a spectrograph (Acton SP2300, 150 grooves/mm grating). Reflectance was calibrated using an aluminium film standard. Data were processed using a custom MATLAB script.

Microscale transmittance measurements were performed using the same microscatterometry setup described above, including the microscope, objective, spectrograph, and CCD detector. Samples were illuminated in transmission mode using a broadband light source, and transmitted light was collected through the objective and projected onto the detector. Although back focal plane images were acquired, intensity was integrated across all collection angles to obtain a single transmittance spectrum. Transmittance was calibrated against an open aperture reference, and data were processed in R (v4.3.2) using standard functions \cite{ref71}.

\subsection{Microstructure observations}

Most of the electron microscopy visualization from this study was carried out at the Ian Holmes Imaging Centre in the Bio21 Molecular Science and Biotechnology Institute of the University of Melbourne.

To quantify the nanoscopic features in the fiddler elytra, we used transmission electron microscopy (TEM). Samples (\textasciitilde{}0.5 $\times$ 0.5 mm) were fixed in 2.5\% glutaraldehyde in PBS and processed for block-face imaging using a standard heavy metal staining protocol. After fixation, samples were stained with 1\% thiocarbohydrazide in Milli-Q water, followed by incubation in 1.5\% potassium ferrocyanide and 2\% OsO\textsubscript{4} in cacodylate buffer. A second osmium staining step was performed using 2\% OsO\textsubscript{4} in cacodylate buffer.  Samples were then incubated overnight at 4$^\circ$C in 1\% aqueous uranyl acetate, followed by 30 minutes at 60$^\circ$C in Walton's lead aspartate. After thorough washing, they were dehydrated through graded ethanol/water and acetone/ethanol series, then infiltrated with graded mixtures of Epon resin and acetone. Finally, samples were embedded in Epon resin with polymerizing agent and cured at 60$^\circ$C for at least 48 hours. We obtained sections of 80 nm thickness with an EM UC7 ultramicrotome (Leica, Germany) and imaged them using a Tecnai Spirit transmission electron microscope (FEI -- ThermoFisher, United States).

We used high resolution scanning electron microscopy (SEM) to study the beetle elytra and photonic structures at different magnifications. We used a sharp minora blade to divide the 1mm diameter circles (from the tissue puncher) into two under a dissecting light microscope. The cut faces were mounted vertically on individual 45$^\circ$ or 90$^\circ$ angled SEM pin stubs using conductive carbon tape. For the inverted contrast images, we instead mounted 200 nm semithin sections cut from the spurs resin blocks were mounted onto a glow-discharged microscope slide, which was then secured to a standard SEM stub using conductive carbon tape. In both cases, samples were sputter coated with a 7nm layer of gold. Samples were imaged using a SU7000 field emission scanning electron microscope (Hitachi, Japan) at 3.00 kV and working distance 4 mm. We used a backscattered electron detector (capturing signal from higher angles and deeper regions of the sample), a secondary electron detector (sensitive to inelastic collisions and surface topography), and an upper detector positioned further from the sample, providing a high-resolution composite of surface and subsurface features. Additional SEM images were acquired on a dual-beam FIB-SEM microscope (model, manufacturer) at Columbia University, using the same sample preparation described above. Images were collected at an accelerating voltage of 1.0 kV using the InLens detector and a working distance of 3.9 mm.

\subsection{Computational models}

We studied the optical properties of the fins using two complementary Finite-Difference Time-Domain (FDTD) simulations (Ansys Lumerical FDTD 2025): 1) an infinitely wide patterned waveguide model and 2) a free-space model. The surface features of the fin were represented as chitin spheres supported by chitin cylinders (clavate lattice) and spherical indentations as air hemispheres (pitted lattice), in perfect triangular lattices. All features in both models were parameterised using the means of measurements extracted from the electron microscopy images.

In the model of an infinitely wide patterned waveguide, we used a single unit cell periodic along the x-axis patterned with surface features. A mode source in two different orthogonal polarisations (TE/TM) was injected directly into the core of the waveguide and monitors recorded reflection and transmission within the core. We simulated both a lossless material model (n=1.55, k=0) and a realistic material model for the refractive index of chitin where the refractive index and absorption coefficient K are wavelength dependent \cite{ref30}.

In the free-space model, we constructed a more realistic representation of a single photonic fin. The core of the fin was represented as a chitin half-elliptic cylinder, with both flat surfaces patterned with the composite triangular lattices. A diffracting plane wave source with highly absorptive (PML) boundary conditions was used. In the simulation, a box of monitors surrounded the source and structure, and far field projections were created from this box of monitors using farfieldexact. These projections were then integrated over selected angular cones to produce the power transmission spectra. Finally, normalized reflectance spectra were obtained by integrating over the full sphere and matching total power transmission to the total Lumerical-normalized transmission through the box of monitors (Fig.~\ref{fig:s12}). Simulated scatterometry images, inspired by the work of Stavenga et. al, and Bauernfeind et. al \cite{ref31,ref32}, were created by converting angularly resolved far field projection spectra into human perceived colour using a custom-written Python script and normalizing by the maximum luminance (Y in XYZ colour space).

\section*{Acknowledgments}

The organisms studied in this research originate from the lands of the Wurundjeri Woi-wurrung and Bunurong Peoples, whose enduring connection to Country we respectfully acknowledge. We thank the Entomological Society of Victoria for help with beetle collection and Dr Chris Moessender insights on cetonid biology; and Nicole Gunter, for facilitating collection access at the Queensland Museum. We thank the team at the Ian Holmes Imaging Centre for microscopy advice and training, and Terry McGlynn and Allison Shultz for access to spectroscopy equipment.

\paragraph{Funding.} Native Australian Animal Trust - Big Science Pitch - University of Melbourne (L.O.R.); Australian Research Council grant DP230100207 (N.Y., D.S-F); Moore Foundation grant GBMF11561 (N.Y.).

\paragraph{Author contributions.}
Conceptualization: L.O.R., N.Y., D.S-F.;
Methodology: L.O.R., N.S.K, Z.Z.;
Investigation: L.O.R., N.S.K, Z.Z., M.W., J.A.H.;
Visualization: L.O.R., N.S.K.;
Supervision: N.Y., D.S-F.;
Writing---original draft: L.O.R., N.S.K.;
Writing---review \& editing: N.Y., D.S-F, J.A.H.
\paragraph{Competing interests.} The authors declare they have no competing interests.

\paragraph{Data and materials availability.} All data and code needed to evaluate and reproduce the results of this manuscript are available on Dryad (\url{https://doi.org/10.5061/dryad.v15dv42cg}). This study did not generate any new materials.

\clearpage
\appendix
\section*{Supplementary Materials}
\addcontentsline{toc}{section}{Supplementary Materials}
\renewcommand{\figurename}{Fig.}
\renewcommand{\thefigure}{S\arabic{figure}}
\setcounter{figure}{0}
\renewcommand{\thetable}{S\arabic{table}}
\setcounter{table}{0}
\noindent This appendix includes Figs.~S1 to S12 and Table~S1.

\begin{figure}[htbp]
\centering
\includegraphics[width=\textwidth,height=0.62\textheight,keepaspectratio]{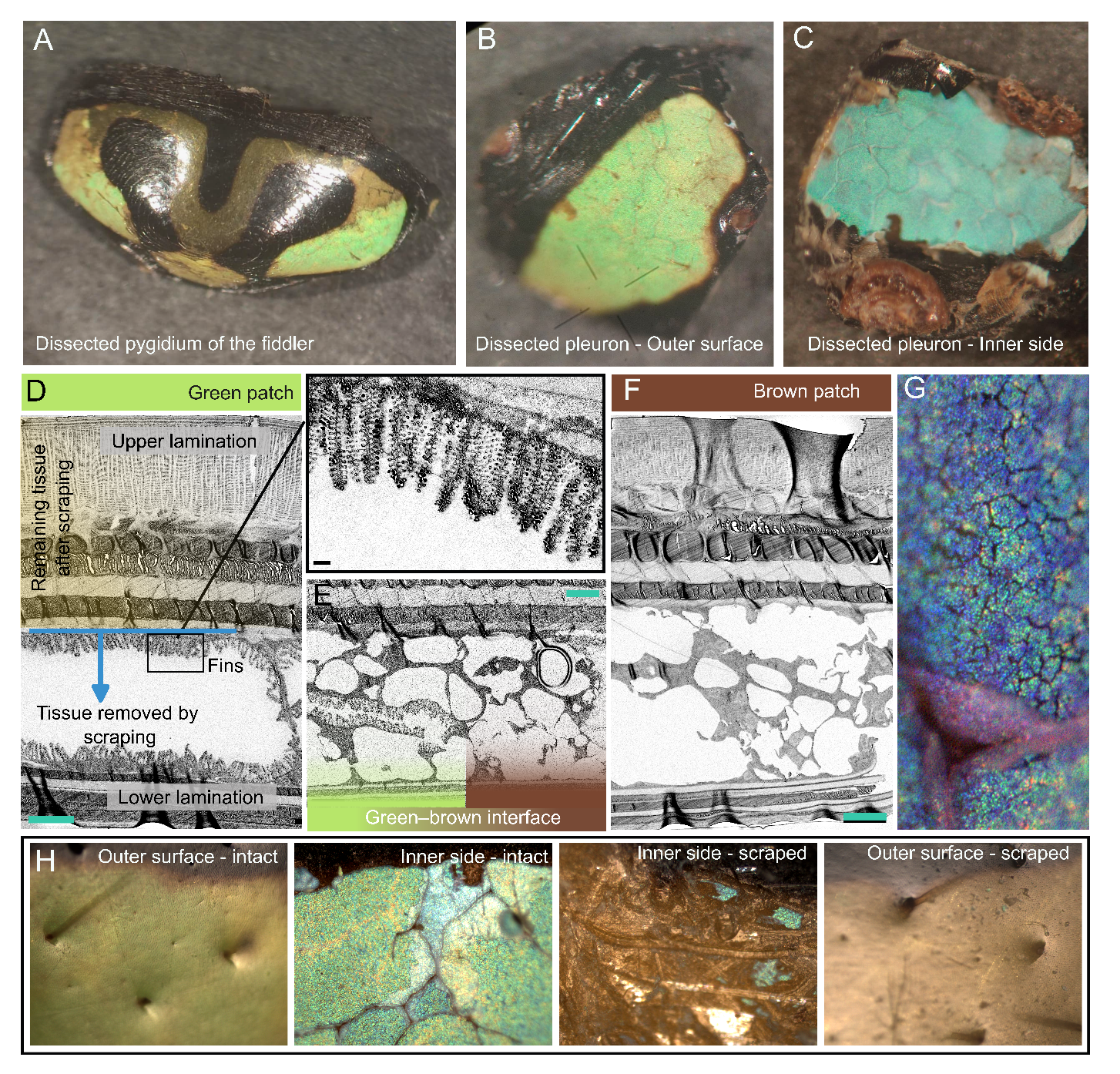}
\caption{\textbf{The green colouration in the fiddler stripes is associated with the presence of fins.} \textbf{(A--C)} Examples of the fiddler beetle stripes dissected and observed under a dissection microscope. \textbf{(D--F)} Inverted contrast SEM images of secondary and low angle backscattered electrons in 200nm sections. \textbf{(D)} Intact green patch in the elytron. The fins are located deep under the surface between the two laminations. In our manual scraping experiments to remove the fins, only the cuticle in the upper lamination remains (highlighted). \textbf{(E)} In patches obtained from the interface between the green stripes and the surrounding brown regions, the fins are only present in the green side. The photonic fins are packed selectively in certain ``vesicles'' of tissue between the two laminations of the elytron. \textbf{(F)} Brown patch, where the fins are absent. \textbf{(G)} light microscopy image of the fins attached in the upper lamination observed from the ventral side. The patchy appearance of this colourful mosaic could be correlated to the fins being contained inside specialized sacs. \textbf{(H)} Scraping experiment performed on a fragment of a green stripe imaged at 20$\times$ magnification. The first panel shows the intact stripe viewed from the outer surface of the elytron. The second panel shows the same fragment viewed from its internal side after careful removal of the lower lamination, exposing the fin layer attached to the upper lamination. The third panel shows the same fragment after the exposed fins were mechanically removed using fine instruments, leaving only the cuticle (shaded region in \textbf{(D)}). The final panel shows this scraped region viewed again from the outer surface.}
\label{fig:s1}
\end{figure}
\clearpage

\begin{figure}[htbp]
\centering
\includegraphics[width=\textwidth,height=0.62\textheight,keepaspectratio]{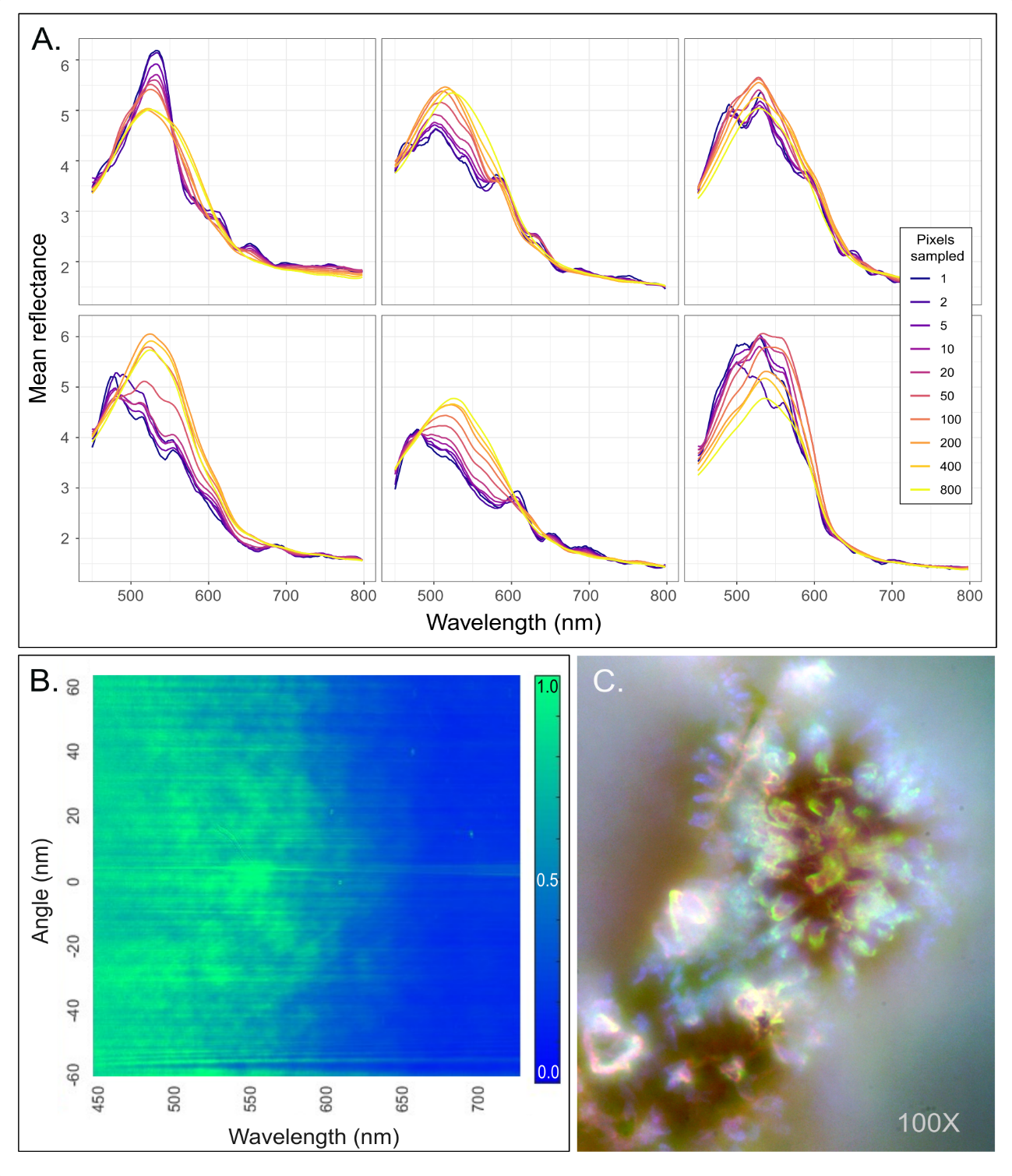}
\caption{\textbf{Spectral variability across hyperspectral imaging pixels and angle-dependent reflectance of the fiddler beetle fin arrays.}
\textbf{(A)} Mean reflectance spectra calculated from increasing numbers of neighbouring pixels within six representative hyperspectral regions of interest. Curves illustrate rapid convergence towards a stable mean spectrum as additional pixels are included. \textbf{(B)} Reflectance measured from a 30um diameter spot on the exposed fins layer, showing the persistence of a diffuse blue--green spectral signal. Reflectance values are normalised to the maximum intensity. \textbf{(C)} Bright-field optical micrograph (100$\times$ objective) of a group of exposed fins over a folded surface in the inner side of the elytron showing qualitatively, that the effective angle of orientation of each fin can modify their reflectance.}
\label{fig:s2}
\end{figure}
\clearpage

\begin{figure}[htbp]
\centering
\includegraphics[width=\textwidth,height=0.62\textheight,keepaspectratio]{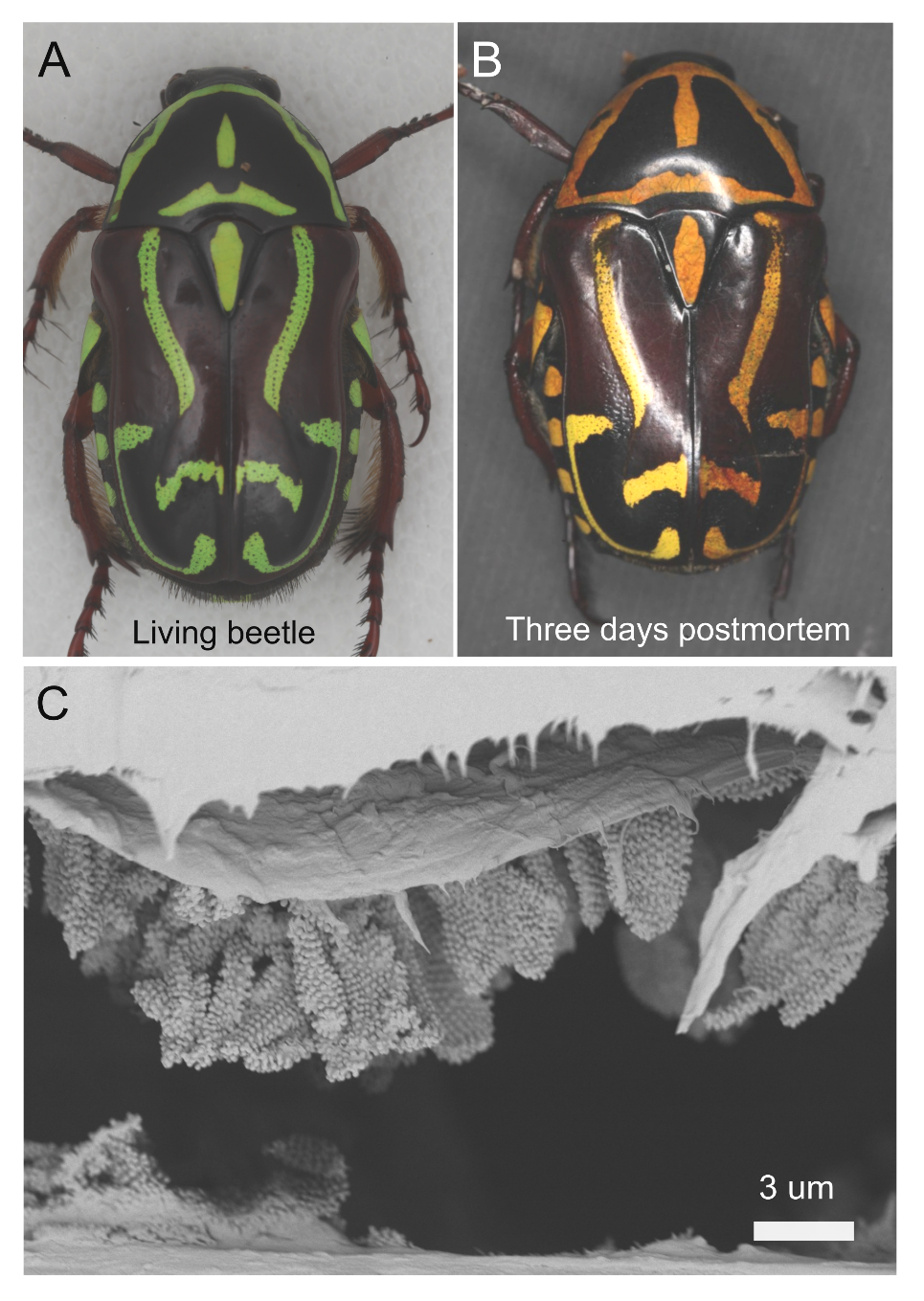}
\caption{\textbf{Structural colour that fades.} \textbf{(A)} A living specimen of \textit{Eupoecila australasiae} displaying the characteristic vivid green stripes. \textbf{(B)} A fiddler specimen three days after death, showing extensive fading of the green colouration. \textbf{(C)} SEM image of a faded stripe revealing that the subsurface fin array remains present despite the loss of green colouration. Although the fins appear distorted relative to those observed in fresh material, the overall architecture is retained. Panels A--C are illustrative and were obtained from different specimens.}
\label{fig:s3}
\end{figure}
\clearpage

\begin{figure}[htbp]
\centering
\includegraphics[width=\textwidth,height=0.62\textheight,keepaspectratio]{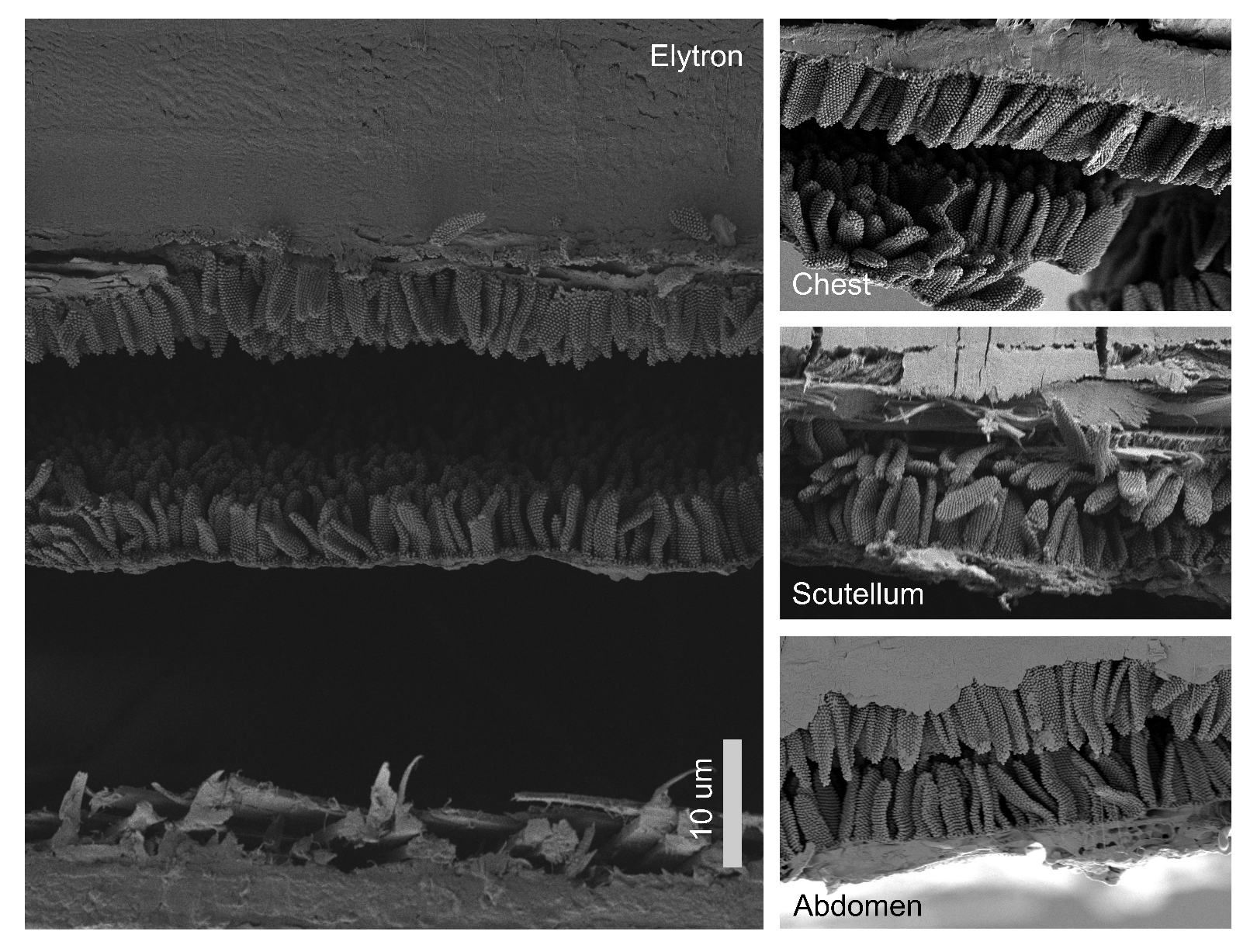}
\caption{\textbf{Alignment of fins within the fin-bearing sacs in different regions of the body.} The spacing between the two opposing rows of fins varies considerably. In some regions, the rows are closely apposed with little or no intervening space, whereas in the elytra they are often separated and the lower row may contact the lower lamination.}
\label{fig:s4}
\end{figure}
\clearpage

\begin{figure}[htbp]
\centering
\includegraphics[width=\textwidth,height=0.62\textheight,keepaspectratio]{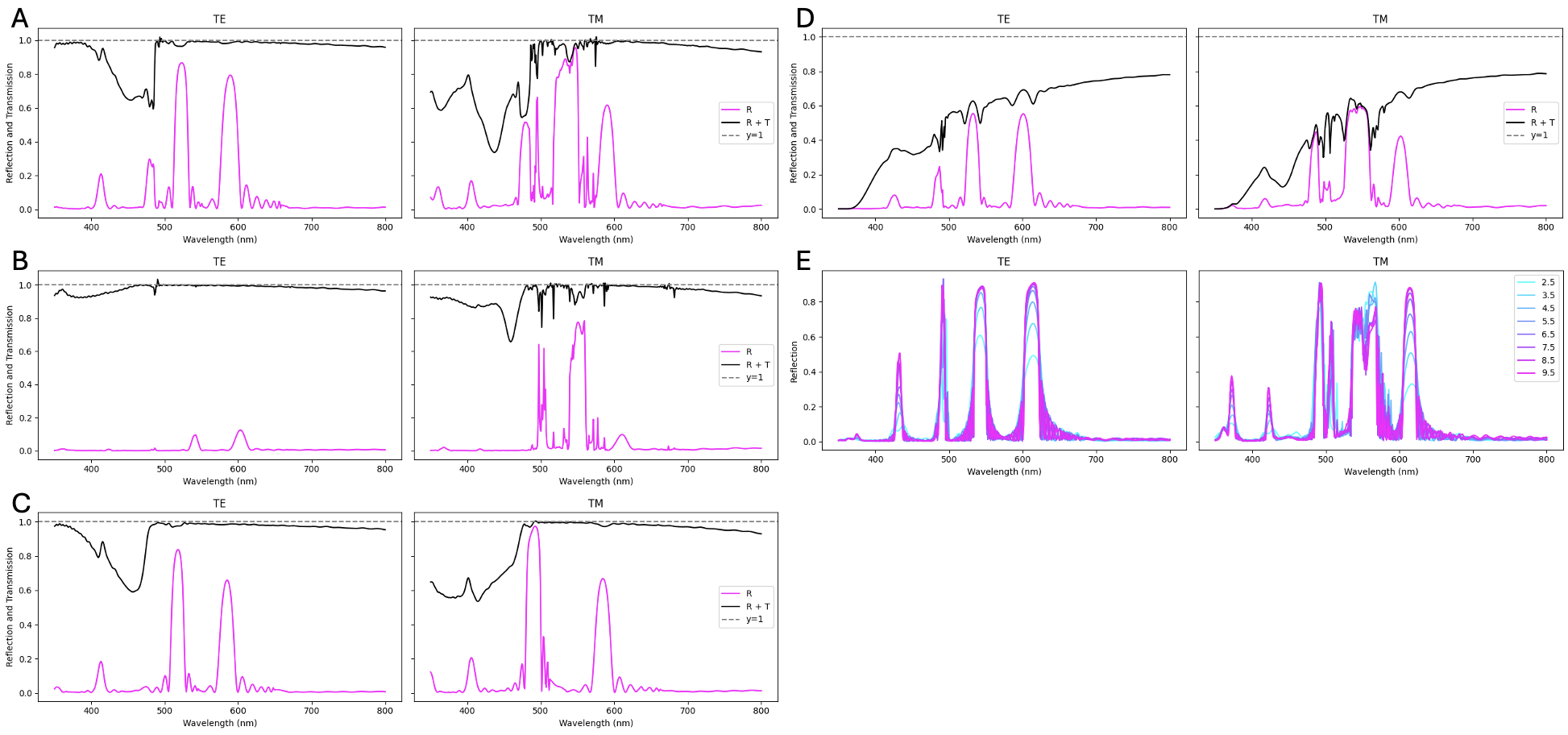}
\caption{\textbf{Stepwise structural simplification in the infinitely wide waveguide model.} Reflection spectra for the infinitely wide waveguide model with \textbf{(A)} all surface features, \textbf{(B)} clavate lattice: the chitin core, chitin cylinders, and chitin spheres (lacking air hemispheres), and \textbf{(C)} pitted lattice: the chitin core and air hemispheres (lacking chitin cylinders and chitin spheres). A lossless model for chitin is used (n=1.55, k=0). We also used a realistic model for chitin (Azofeifa et al., 2012) for \textbf{(D)} all surface features and \textbf{(E)} all surface features, varying the length of the region with surface features. The mean true length of the fins is close to 4.5 microns, which, as shown in \textbf{(E)}, maintains a high reflection peak while keeping the fin length short, possibly for mass conservation. Mode source analyses show that clavate surface features (spheres and cylinders) and pitted surface features (spherical indentations) each have roles for specular reflection. Clavate features are particularly important in inciting reflection with the TM mode, whereas pitted features are important for both TE and TM modes.}
\label{fig:s5}
\end{figure}
\clearpage

\begin{figure}[htbp]
\centering
\includegraphics[width=\textwidth,height=0.62\textheight,keepaspectratio]{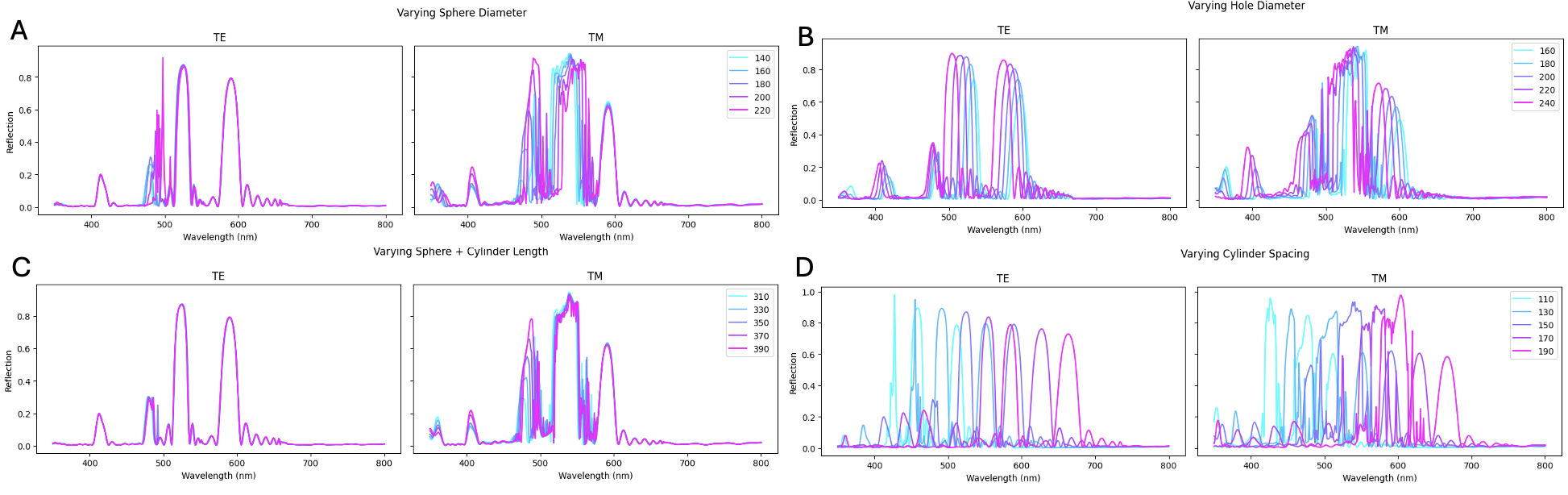}
\caption{\textbf{Sensitivity analysis in the infinitely wide waveguide model.} Reflection spectra for the infinitely wide waveguide model with variations in surface feature dimensions. \textbf{(A)} Varying the chitin sphere diameter. \textbf{(B)} Varying the air hemisphere diameter. \textbf{(C)} Varying the cylinder length. Cylinder + sphere length is reported in the legend. \textbf{(D)} Varying the lattice constant of the triangular lattice of surface features. Cylinder spacing, as reported in the legend, refers to the air gap between cylinders, i.e. lattice constant -- cylinder diameter. Reflection peak locations are more sensitive to the triangular lattice constant dimension, over any of the other tested dimensions of the surface features.}
\label{fig:s6}
\end{figure}
\clearpage

\begin{figure}[htbp]
\centering
\includegraphics[width=\textwidth,height=0.62\textheight,keepaspectratio]{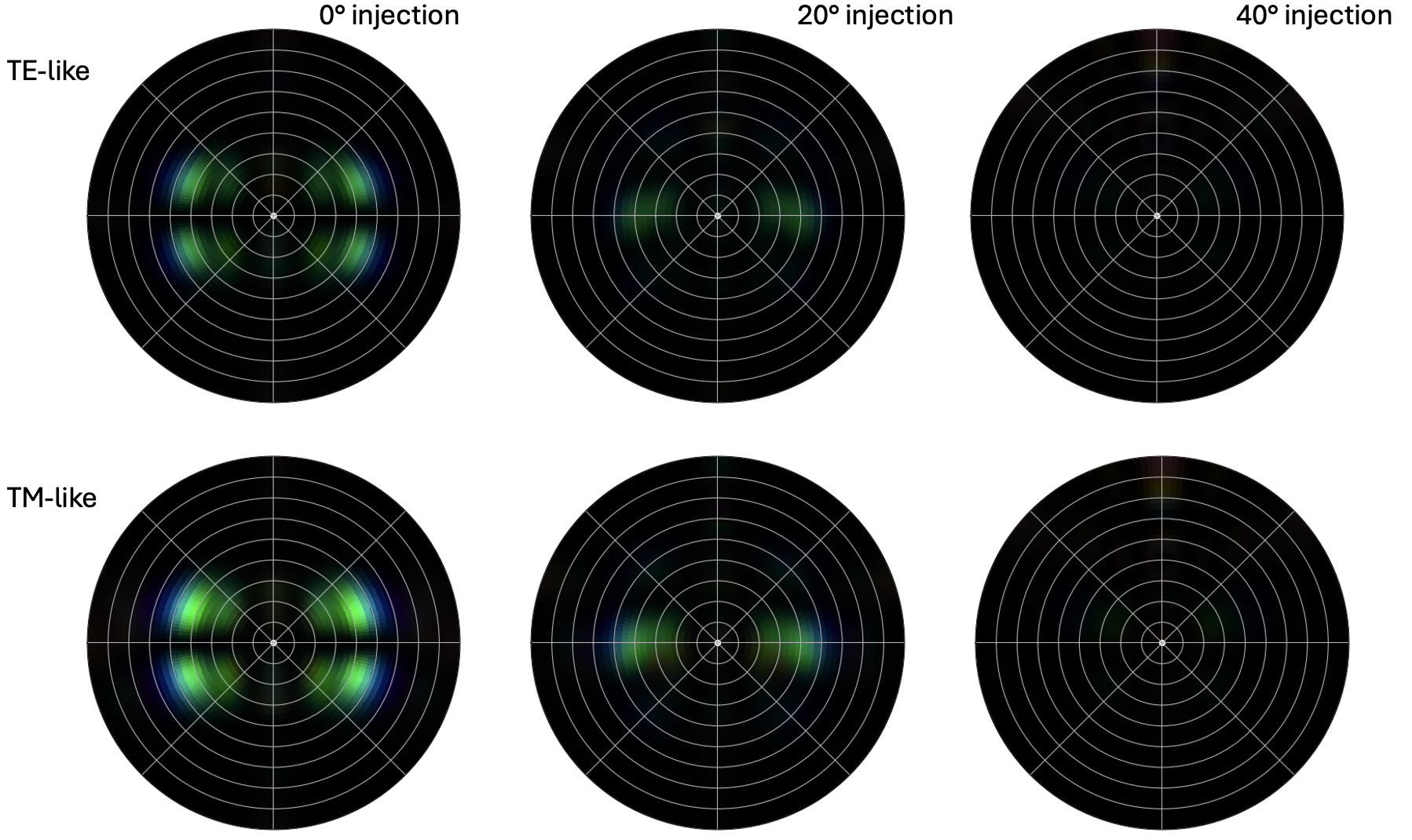}
\caption{\textbf{Scatterometry images of angled injection for two photonic crystal slabs floating in air (no core).} Compared to Figure 4 in the main text, the central core is integral in supporting guided mode backscattering the reddish wavelengths) as well as stabilizing the strength of reflection with increased source injection angle.}
\label{fig:s7}
\end{figure}
\clearpage

\begin{figure}[htbp]
\centering
\includegraphics[width=\textwidth,height=0.62\textheight,keepaspectratio]{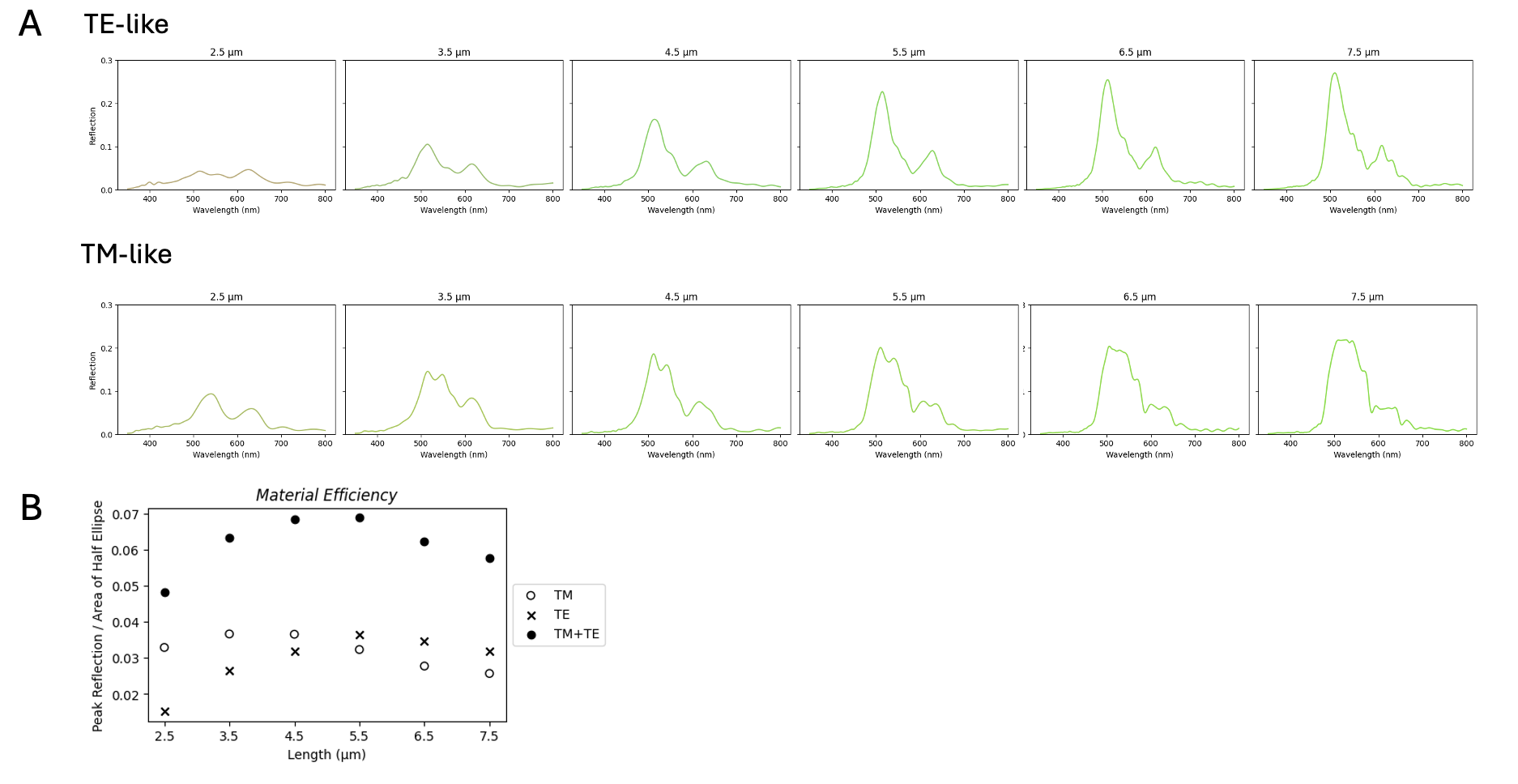}
\caption{\textbf{Effect of the fin length.} \textbf{(A)} Reflection spectra for the free-space model with various fin lengths for TE-like and TM-like excitation, displayed as 45$^\circ$ integrated back scattering. The closest length to the mean measured fin length is 4.5$\mu$m. \textbf{(B)} We investigated a figure of merit termed ``material efficiency'', defined here as the ratio of peak back-scattering and estimated fin mass (which we approximate as the area of the half-ellipse) for TE-like and TM-like excitation, as well as their sum. The mean measured fin length (\textasciitilde{}4.5$\mu$m) is near-optimal in this estimation.}
\label{fig:s8}
\end{figure}
\clearpage

\begin{figure}[htbp]
\centering
\includegraphics[width=\textwidth,height=0.62\textheight,keepaspectratio]{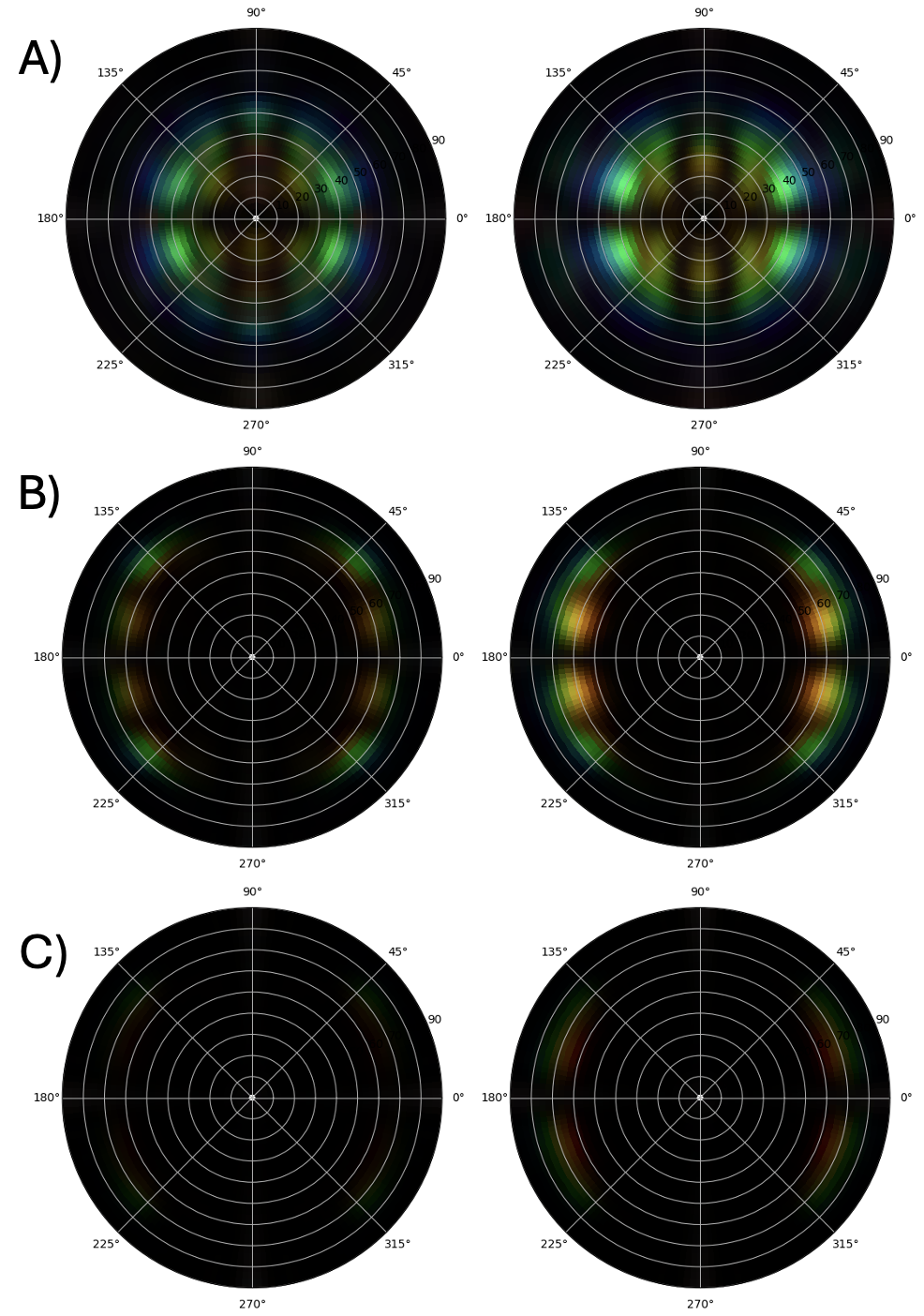}
\caption{\textbf{Effects of changes in refractive index contrast.} Simulated scatterometry images for a fin in refractive index \textbf{(a)} n=1 (air), \textbf{(b)} n=1.34 (like hemolymph), and \textbf{(c)} n=1.47 (like cantharidin oil) for TE-like and TM-like excitation. Intensity scaled by greatest luminance across all images (occurs in TM-like air simulation). This could be a reason for changes in colour of the fiddler beetle stripes, post-mortem.}
\label{fig:s9}
\end{figure}
\clearpage

\begin{figure}[htbp]
\centering
\includegraphics[width=\textwidth,height=0.62\textheight,keepaspectratio]{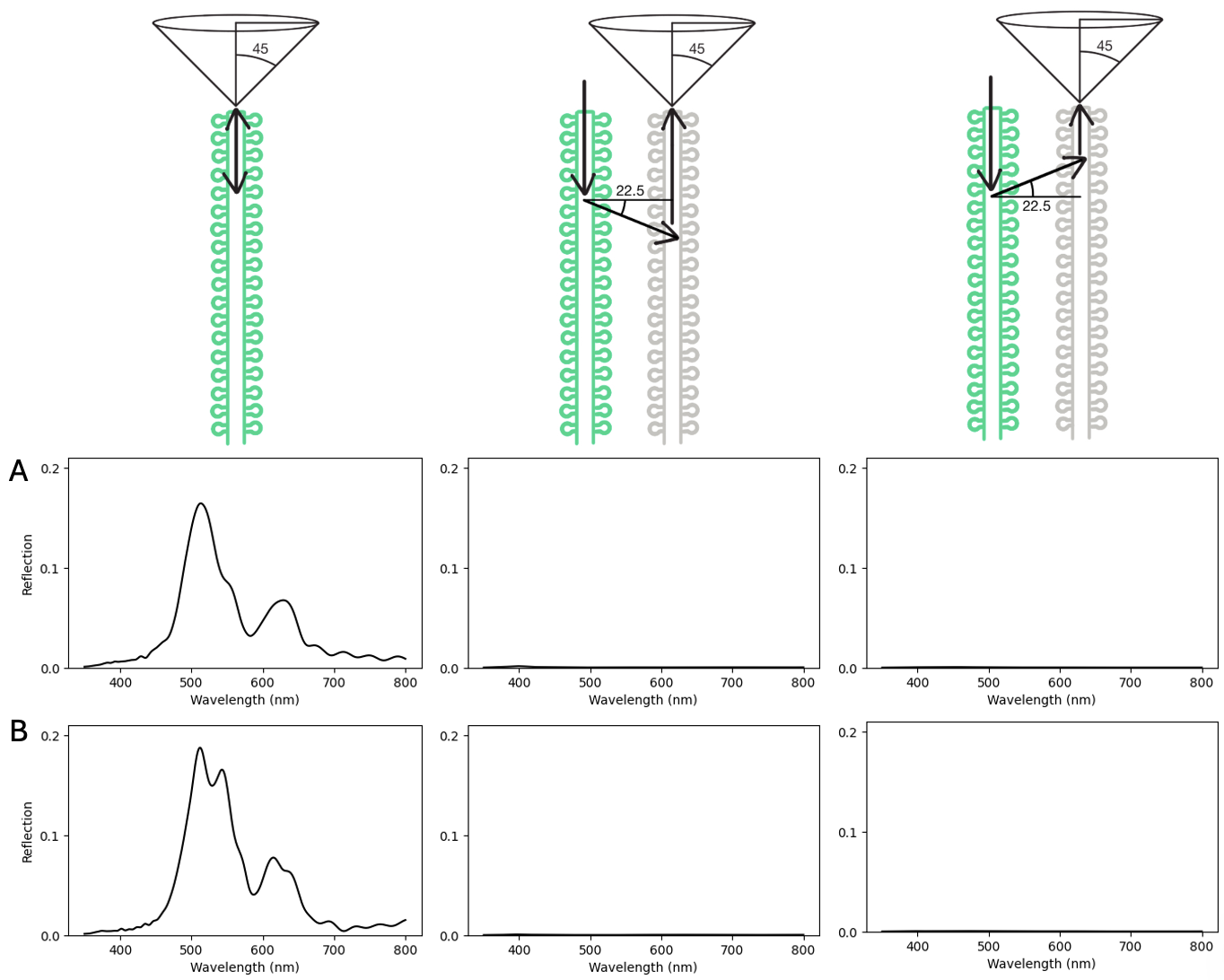}
\caption{\textbf{Effects of secondary scattering.} The effects of light scattered by one fin, entering another, and being transmitted back to the viewer, is on the order of 1/100th of direct reflection. Direct reflection is represented as normalized power transmission (reflectance) through an angular cone of half angle 45$^\circ$ centered on the z+ axis (column 1). Secondary scattering is approximated by multiplying a fin's reflectance between 45-90$^\circ$ by the reflectance through a 45$^\circ$ angular cone for a fin model with the source injected from the side at 22.5$^\circ$ (columns 2 and 3). Spectra are normalized as described in Methods. Note that this calculation ignores interference. \textbf{(A)} TE-like excitation. \textbf{(B)} TM-like excitation.}
\label{fig:s10}
\end{figure}
\clearpage

\begin{figure}[htbp]
\centering
\includegraphics[width=\textwidth,height=0.62\textheight,keepaspectratio]{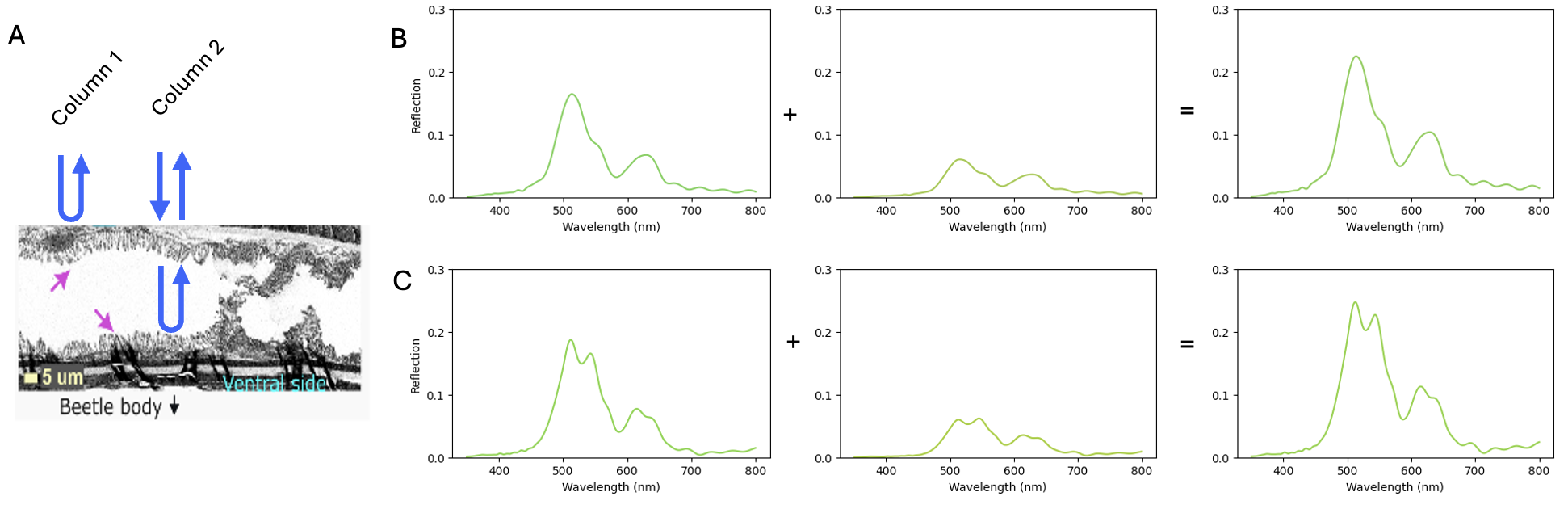}
\caption{\textbf{A double layer of fins enhances, but does not shift, the reflection peaks.} The reflectance of one layer of fins (R) is summed with the reflectance of the second layer of fins (T*R*T) to produce the final reflectance spectrum. Spectra are normalized as described in Methods. Note that this calculation ignores interference. \textbf{(A)} Schematic of reflection from two layers of fins, indicated by blue arrows. \textbf{(B)} TE-like excitation. \textbf{(C)} TM-like excitation.}
\label{fig:s11}
\end{figure}
\clearpage

\begin{figure}[htbp]
\centering
\includegraphics[width=\textwidth,height=0.62\textheight,keepaspectratio]{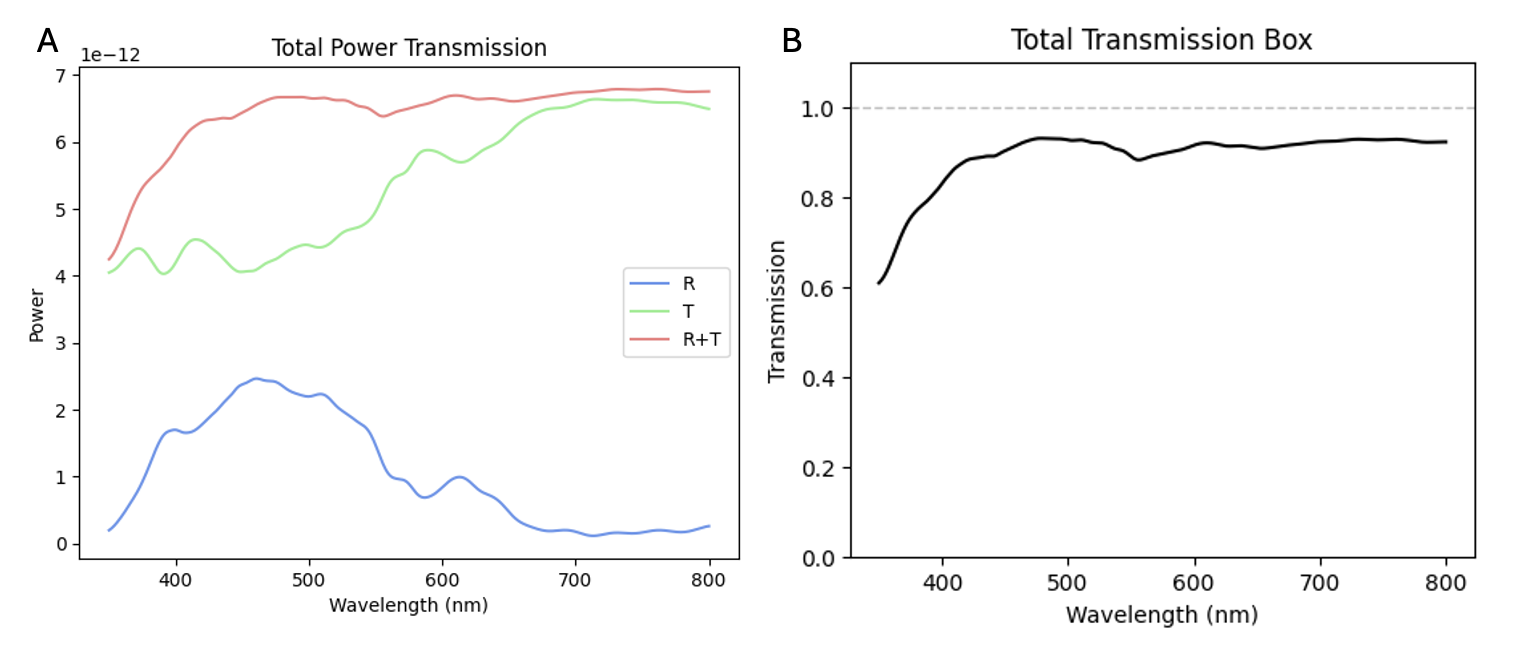}
\caption{\textbf{Normalization of far field power spectra.} Far field power transmission spectra were normalized for each free-space simulation. Here we provide an example simulation with \textbf{(A)} integration over the z+ hemisphere (R, blue), integration over the z- hemisphere (T, green), their sum, i.e. integration over all space (R+T, red) and \textbf{(B)} sum of Lumerical-normalized transmission through the box of monitors surrounding the free-space fin model. Note how the curve shape for total power transmission and total Lumerical-normalized transmission are nearly identical, allowing the normalization of power transmission curves for all selected angular cones.}
\label{fig:s12}
\end{figure}
\clearpage

\begin{table}[htbp]
\centering
\footnotesize
\captionsetup{labelformat=default}
\caption{Dimensions of the different features in the photonic fins.}
\label{tab:s1}
\begin{tabular}{p{2.8cm}ccccccp{2.8cm}}
\toprule
measurement & n & mean & sd & median & min & max & Source \\
\midrule
Fin length ($\mu$m) & 26 & 4.59 & 0.854 & 4.67 & 2.51 & 6.02 & 8 per beetle \\
Fin's area ($\mu$m$^2$) & 5 & 6.62 & - & 6.76 & - & - & Only the best from TEM images \\
Fin's max Ferret & 5 & 5.02 & - & 5.26 & - & - & Only the best from TEM images \\
Fin's min Ferret & 5 & 1.71 & - & 1.66 & - & - & Only the best from TEM images \\
Pit diameter (nm) & 142 & 195 & 37.6 & 197 & 108 & 331 & 90 per beetle \\
Sphere diameter (nm) & 280 & 176 & 23.6 & 177 & 116 & 243 & 90 per beetle \\
Sphere spacing (nm) & 50 & 105 & 34.9 & 112 & 25.9 & 160 & 3 plot profile per beetle. 10 to 27 spaces per plot profile \\
Rod spacing (nm) & 50 & 144 & 49.4 & 149 & 0 & 222 & 3 plot profile per beetle. 10 to 27 spaces per plot profile \\
Rod + sphere length (nm) & 90 & 349 & 54.8 & 344 & 237 & 464 & 30 per beetle \\
Calculated rod length (nm) &  & 173.6 & - & 167 & - & - & (Rod + sphere length) -- (sphere diameter) \\
\bottomrule\end{tabular}\end{table}

\end{document}